\documentclass[reqno,12pt]{article}
\usepackage{arydshln}
\usepackage[usenames,dvipsnames]{xcolor}
\usepackage{ae}
\usepackage[T1]{fontenc}
\usepackage[ansinew]{inputenc}
\usepackage{mathrsfs}
\usepackage{amsmath}
\usepackage{amssymb}
\usepackage{graphicx}
\usepackage{color}
\definecolor{darkblue}{cmyk}{0.9,0.9,0,0}
\definecolor{carageen}{RGB}{0,0.55,0}
\usepackage[colorlinks=true,linkcolor=darkblue,citecolor=darkblue,urlcolor=darkblue]{hyperref}
\usepackage{epsfig}
\usepackage{cite}

\newcommand{\M}{{\cal M}}

\newcommand{\comment}[1]{}

\newcommand{\beq}{\begin{equation}}
\newcommand{\eeq}{\end{equation}}
\newcommand{\beqq}{\begin{equation*}}
\newcommand{\eeqq}{\end{equation*}}
\newcommand\beqa{\begin{eqnarray}}
\newcommand\eeqa{\end{eqnarray}}
\newcommand\beqaa{\begin{eqnarray*}}
\newcommand\eeqaa{\end{eqnarray*}}
\newcommand\bea{\begin{array}}
\newcommand\eea{\end{array}}

\def\Xint#1{\mathchoice
{\XXint\displaystyle\textstyle{#1}}
{\XXint\textstyle\scriptstyle{#1}}
{\XXint\scriptstyle\scriptscriptstyle{#1}}
{\XXint\scriptscriptstyle\scriptscriptstyle{#1}}
\!\int}
\def\XXint#1#2#3{{\setbox0=\hbox{$#1{#2#3}{\int}$ }
\vcenter{\hbox{$#2#3$ }}\kern-.5\wd0}}

\def\dashint{\Xint-}

\newcommand\IM{{\rm Im}\,}
\newcommand\RE{{\rm Re}\,}

\def\Xint#1{\mathchoice
{\XXint\displaystyle\textstyle{#1}}
{\XXint\textstyle\scriptstyle{#1}}
{\XXint\scriptstyle\scriptscriptstyle{#1}}
{\XXint\scriptscriptstyle\scriptscriptstyle{#1}}
\!\int}
\def\XXint#1#2#3{{\setbox0=\hbox{$#1{#2#3}{\int}$}
\vcenter{\hbox{$#2#3$}}\kern-.5\wd0}}

\newcommand{\nn}{\nonumber}

\newcommand{\neqa}{\nonumber\end{eqnarray}}
\newcommand{\la}[1]{\label{#1}}

\newcommand{\eq}[1]{(\ref{#1})}

\newcommand{\T}{{\cal T}}

\newcommand{\hs}{\frac{\sqrt{3}}{2}}
\renewcommand{\d}{\partial}

\newcommand{\<}{{\langle}}
\renewcommand{\>}{{\rangle}}

\newcommand{\re}{\relax{\rm I\kern-.18em R}}

\renewcommand{\sp}{p\hspace{-.40em}/}

\def\su2{{SA(2)}}

\def\[{\left[}
\def\]{\right]}

\def\s{\sigma}

\def\({\left(}
\def\){\right)}
\def\[{\left[}
\def\]{\right]}

\def\<{\langle}
\def\>{\rangle}

\def\bT{{\bf T}}

\def\cY{{\cal Y}}
\def\cX{{\cal X}}

\def\mC{{\mathbb C}}

\def\s*{\ *_{\!\!\!\!\!\!\!\!\!\,_{\,_\text{\scriptsize{sys}}}}}
\def\hs*{\ \hat{*}_{\!\!\!\!\!\!\!\!\!\,_{\,_\text{\scriptsize{sys}}}}}
\def\d{\partial}

\def\sK{\,\slash\!\!\!\! K}

\def\i2{\frac{i}{2}}

\def\bQ{{\bf Q}}

\def\spi{\relax{\rm \pi\kern-0.5em /}}
\def\sA{\relax{\rm A\kern-0.5em /}}
\def\sp{\relax{\rm p\kern-0.5em /}}
\def\sd{\relax{\rm \d\kern-0.5em /}}
\def\sk{\relax{\rm k\kern-0.5em /}}
\def\sn{\relax{\rm n\kern-0.5em /}}
\def\sl{\relax{\rm l\kern-0.5em /}}
\def\sP{\relax{\rm P\kern-0.7em /}}
\def\sBethe{\relax{\rm \Bethe\kern-0.5em /}}

\def\cT{{\cal T}}
\def\cY{{\cal Y}}

\newcommand{\Blue}[1]{{\color{blue}#1\color{black}}}

\usepackage{varioref}
\usepackage{makeidx}
\makeindex

\usepackage[english]{babel}
\begin{document}
\thispagestyle{empty}

\renewcommand{\thefootnote}{\fnsymbol{footnote}}

\vfill
\begin{flushright}
\it Dedicated to our dear friend Pedro birthday
\end{flushright}

\setcounter{page}{1}
\setcounter{footnote}{0}
\setcounter{figure}{0}
\begin{center}
$$$$
{\Large\textbf{\mathversion{bold}
Analytic Solution of Bremsstrahlung TBA
}\par}

\vspace{1.0cm}

\textrm{Nikolay Gromov$^{\,\phi}$ and Amit Sever$^{\,\theta}$}
\\ \vspace{1.2cm}
\footnotesize{
\textit{$^{\phi}$ King's College London, Department of Mathematics WC2R 2LS, UK}\\
 \& \\
\textit{St.Petersburg INP, St.Petersburg, Russia} \\
\texttt{nikgromov@gmail.com} \\
\vspace{5mm}
\textit{$^{\theta}$ Perimeter Institute for Theoretical Physics\\ Waterloo,
Ontario N2L 2Y5, Canada}  \\
\&\\
\textit{School of Natural Sciences,\\Institute for Advanced Study, Princeton, NJ 08540, USA} \\
\texttt{amit.sever@gmail.com}
\texttt{}
\vspace{3mm}
}

\par\vspace{1.5cm}

\textbf{Abstract}\vspace{2mm}
\end{center}

We consider the quark--anti-quark potential on the three sphere or the generalized cusp anomalous dimension in planar ${\cal N}=4$ SYM. We concentrate on the vacuum potential in the near BPS limit with $L$ units of R-charge. Equivalently, we study the anomalous dimension of a super-Wilson loop with $L$ local fields inserted at a cusp. The system is described by a recently proposed infinite set of non-linear integral equations of the
Thermodynamic Bethe Ansatz (TBA) type. That system of TBA equations is very similar to the one of the spectral problem but simplifies a bit in the near BPS limit. Using
techniques based on the Y-system of functional equations we first reduced the infinite system of TBA
equations to a Finite set of Nonlinear Integral Equations (FiNLIE).
Then we solve the FiNLIE system analytically, obtaining a simple analytic result for the potential!
Surprisingly, we find that the system has  equivalent descriptions in terms of an
effective Baxter equation and in terms of a matrix model.
At $L=0$, our result matches  the one obtained before using localization techniques.
At all other $L$'s, the result is new. Having a new parameter, $L$, allows us to take the large $L$ classical limit.
We use the matrix model
 description to solve the classical limit and match the result with a string theory computation. Moreover, we find that the classical string algebraic curve matches
 the algebraic curve arising from the matrix model.

\noindent

\vfill
\begin{flushright}
\end{flushright}

\setcounter{page}{1}
\renewcommand{\thefootnote}{\arabic{footnote}}
\setcounter{footnote}{0}

 \def\nref#1{{(\ref{#1})}}

\tableofcontents
\section{Introduction}

Recent developments in quantum field theory suggest that for the first time we may be able to solve an interacting gauge theory in four dimensions. Most of these developments have come about in the study of ${\cal N}=4$ SYM in its planar limit. Planar observables in that theory are mapped to two dimensional problems that are solved exactly using Integrability methods\footnote{The integrability in QCD was first discovered in \cite{Lipatov:1993yb,Faddeev:1994zg}. Very fast development of the integrability in ${\cal N}=4$ SYM
started after the seminal paper \cite{Minahan:2002ve}.}. Such observables include the spectrum, Wilson loops, scattering amplitudes and
correlation functions. Among these, the most developed study so far is of the quantum spectrum of the theory
\cite{Beisert:2010jr}. It is now available in a completely non-perturbative fashion numerically or even analytically in some cases
\cite{Beisert:2006ez,Gromov:2009tv,Bombardelli:2009ns,Gromov:2009bc,Arutyunov:2009ur,Gromov:2009zb}.

Nevertheless, the solution of the spectral problem remains to be rather involved technically \cite{Gromov:2009tv,Bombardelli:2009ns,Gromov:2009bc,Arutyunov:2009ur}
and even sometimes scary \cite{Gromov:2010km,Arutyunov:2011mk}. It is given by an infinite set of functional equations (Y-system) supplemented with definite analytical properties
\cite{Cavaglia:2010nm,Gromov:2011cx}.

Recently, the methods developed for the spectrum of the theory on $S^3$ were extended to a new class of observables -- the spectrum of the color flux between external quarks on $S^3$ \cite{Correa:2012hh,Drukker:2012de}. In its vacuum, the energy of that flux is the same as the generalized cusp anomalous dimension $\Gamma_\text{cusp}$. That is, the conformal dimension of a quark and anti-quark lines meeting at a cusp \cite{PolyakovCusp}
\beq
\<W\>=\({\Lambda_\text{IR}\over\Lambda_\text{UV}}\)^{\Gamma_\text{cusp}}
\eeq
Here $W$ is a cusped Wilson loop representing the quark--anti-quark lines and $\Lambda_\text{UV}$, $\Lambda_\text{IR}$ are the UV and IR cutoffs.\footnote{The phrase ``cusp anomalous dimension" is also used to describe the IR divergence of massless particles. That cusp anomalous dimension $\Gamma_{cusp}^\infty$ is one of the first observables computed exactly using integrability~\cite{Beisert:2006ez}. It controls the behavior of $\Gamma_\text{cusp}$ in the limit where the two quarks are infinitely boosted with respect to each other $\lim\limits_{i\phi\to\infty}\Gamma_\text{cusp}=i\phi/4\times \Gamma_\text{cusp}^\infty$. We hope that this will not cause confusion.} The solution takes the form of a Thermodynamic Bethe Ansatz (TBA) system very similar to the one for the spectrum.
These are an infinite set of non-linear  integral equations.
\begin{figure}[h]
\centering
\def\svgwidth{14cm}
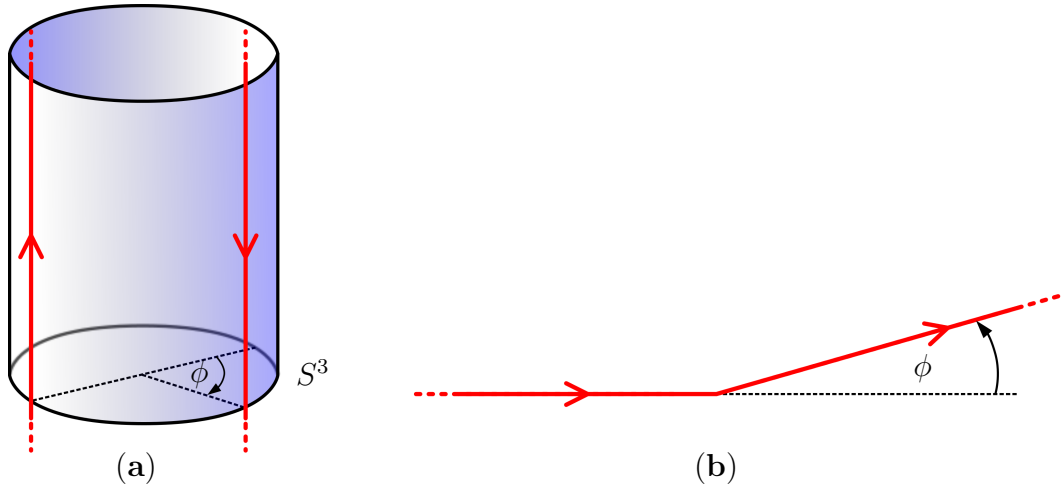
\caption{\it ({\bf a}) A quark--anti-quark pair sitting at two points on $S^3$ at a relative angle $\pi -\phi$. The quark--anti-quark lines are extended along the (Euclidian) time direction. ({\bf b}) Under the cylinder to plane conformal map, the quark and anti-quark lines in ({\bf a}) are mapped to the two half lines of a Wilson line with a cusp angle $\phi$.}
\label{CuspDiagram}
\end{figure}

The most natural quarks in ${\cal N}=4$ SYM are infinitely massive W-bosons on the boundary of the Coulomb branch. These are locally supersymmetric quarks probes that also couple to a scalar. As we have six scalars, this coupling selects a point $\vec n$ in $S^5$. As a result, one of the new key features of the cusp TBA system is that it is parameterized by two continuous parameters. That is, $\Gamma_\text{cusp}$ is a function of
two angles $\phi$ and $\theta$ \cite{Drukker:1999zq}. The angle $\phi$ is the geometrical angle between the two lines, see figure \ref{CuspDiagram}.
The second angle $\theta$, is the angle on $S^5$ between the quark and anti-quark points $\cos\theta=\vec n_q\cdot\vec n_{\bar q}$. The corresponding cusped Wilson loop is 
\beq
W_0={\rm P}\exp\!\int\limits_{-\infty}^0\! dt\[i  A\cdot\dot{x}_q+\vec\Phi\cdot\vec n_q\,|\dot x_q|\]\times {\rm P}\exp\!\int\limits_0^\infty\!dt\[i A\cdot\dot x_{\bar q}+\vec\Phi\cdot\vec n_{\bar q}\,|\dot x_{\bar q}|\]
\eeq
where $\vec\Phi$ is a vectors made of the six scalars of ${\cal N}=4$ SYM. Here, $x_q(t)$ and $x_{\bar q}(t)$ are two straight lines representing the quark and anti-quark trajectories. They connect the origin and infinity such that $\dot x_q\cdot\dot x_{\bar q}/(|\dot x_q||\dot x_{\bar q}|)=\cos\phi$.

The existence of these two parameters allows one to take various limits where the TBA equations simplify. We hope that a better handle on the solution in such limits will shed light on how to simplify the TBA equations in general. One such limit is $\phi^2-\theta^2\to0$. When $\phi^2=\theta^2$ the cusped Wilson loop is BPS and the energy vanishes \cite{Zarembo:2002an}.
Consider the case $\theta=0$.  As we deform the angles away from this supersymmetric configuration, the energy behaves as
\beq \la{DevBPS}
\Gamma_\text{cusp}(\phi,0)  = -\phi^2 B_0(g) + {\cal O}\(\phi^4\)\;.
\eeq
The function $B_0$, also known as the ``Bremsstrahlung function", is related to a variety of physical quantities
\cite{Correa:2012at,Fiol:2012sg}. In particular, it controls the power radiation of an accelerating quark. That function was computed exactly in \cite{Correa:2012at,Fiol:2012sg} using results from localization \cite{localizations,Pestun}.

In the planar limit the localization result gives
\beq
B_0
=g^2\(1-\frac{I_3(4\pi g)  }{I_1(4\pi g) }\)   \la{bplanar}
\eeq
where $I_n$ are modified Bessel functions. On the other hand, the same function $B_0$ is also computed by the TBA. In the near BPS limit the TBA reduces to a somewhat simplified set of equations \cite{Correa:2012hh}, as we review in the next
 section. These are still an infinite set of non-linear integral equations. The simplicity of (\ref{bplanar}) strongly suggests that, in the near BPS limit, the TBA equations can be drastically simplified.
Indeed, we managed to solve the TBA equations analytically in this limit and derive \eq{bplanar}.

Furthermore, one can consider a more general configuration with adjoint fields inserted at the cusp.\footnote{In particular, the TBA equations where derived by first considering the problem with infinite fields insertions at the cusp.} In particular, one can insert $L$ complex scalars $Z=\Phi_1+i\Phi_2$ at the cusp 
\beq
W_L={\rm P}\exp\!\int\limits_{-\infty}^0\! dt\(i  A\cdot\dot{x}_q+\vec\Phi\cdot\vec n\,|\dot x_q|\)\times Z^L\times {\rm P}\exp\!\int\limits_0^\infty\!dt\(i A\cdot\dot x_{\bar q}+\vec\Phi\cdot\vec n\,|\dot x_{\bar q}|\)
\eeq
where $\Phi_1$ and $\Phi_2$ are two scalars different from the one that is coupled to the Wilson lines, ($\vec n\cdot\vec\Phi$). That is a convenient choice for two reasons. First, $W_L$ has a good anomalous dimension. That is because the Wilson lines cannot mix with the $Z$'s and there is no other insertion with the same free charges. Second, at 
$\phi=0$, that configuration preserves supersymmetry \cite{DrukkerKawamoto}. 
This freedom allows us to generalize the Bremsstrahlung function $B_0$ to the case with $Z^L$ insertion -- $B_L$. We solve the TBA for $B_L$ analytically at any $L$! The result is again a rational
function of modified Bessel functions $I_n(4\pi g)$:
\beq\la{BLsimplified0}
B_L=g^2\(-R_{L+1}+2R_L-R_{L-1}\)\;,
\eeq
where\footnote{The dashed line shows where a row was deleted.}
\beq\la{eq5}
R_L\equiv
\left|
\bea{ccccc}
I_{1}&I_{3}&\dots&I_{2L-1}&I_{2L+1}\\ \hdashline
I_{-3}&I_{-1}&\dots&I_{2L-5}&I_{2L-3}\\
\vdots&\vdots& \ddots & \vdots& \vdots\\
I_{1-2L}&I_{3-2L}&\dots&I_{-1}&I_{1}\\
I_{-1-2L}&I_{1-2L}&\dots&I_{-3}&I_{-1}
\eea
\right|/
\left|
\bea{ccccc}
I_{-1}&I_{1}&\dots&I_{2L-3}&I_{2L-1}\\
I_{-3}&I_{-1}&\dots&I_{2L-5}&I_{2L-3}\\
\vdots&\vdots& \ddots & \vdots& \vdots\\
I_{1-2L}&I_{3-2L}&\dots&I_{-1}&I_{1}\\
I_{-1-2L}&I_{1-2L}&\dots&I_{-3}&I_{-1}
\eea
\right|\;.
\eeq

The paper is organized as follows. In section~\ref{sec:red} we first apply the methods of \cite{talk,Gromov:2011cx}\footnote{For a more recent alternative  approach see also \cite{seealso,seealso2}.} to reduce this infinite set of equations
to just three simple integral equations. Then, in section~\ref{sec:ana}, we solve these three equations, obtaining an exact analytic result at any value of the 't Hooft coupling, therefore establishing the drastic simplification of the TBA system! The result of course agrees precisely with (\ref{bplanar})
\cite{Correa:2012at,Fiol:2012sg}. In section~\ref{sec:ana} we analytically solve the TBA system with $L$ units of R-charge, obtaining our main result (\ref{BLsimplified0}). Very likely, the same localization techniques used to derive (\ref{bplanar}) \cite{localizations,Pestun,Correa:2012at,Fiol:2012sg}, can be extended to this case. Therefore, our results provide for the first time predictions for a localization computation from integrability.
In section~\ref{sec:clas} we consider the classical large $L$ limit of the construction. To take this limit we were forced to reformulate our result in terms
of a matrix model partition function which we solve in the classical (thermodynamic) limit. The result is shown to match the classical string prediction
derived in appendix~\ref{app:class}.

\section{Bremsstrahlung TBA}
\la{sec:BremsstrahlungTBA}

In this section we review the cusp TBA system and its reduction in the near BPS limit $\phi^2-\theta^2\to0$, namely the {\it Bremsstrahlung TBA} \cite{Correa:2012hh}. We also use it to introduce our notations. For simplicity, we will restrict our discussion to the case where $\theta=0$ and therefore $\phi$ is the smallest parameter.\footnote{The general near BPS case, with $\theta\ne0$ has a similar degree of complexity. However, we leave that generalization to a future work.}  A reader familiar with \cite{Correa:2012hh} may jump to the next section.

The TBA equations for the spectrum of the color flux between a quark and an anti-quark are very similar to the ones describing the spectrum of single trace operators. I.e. they follow the pattern of the T-hook with three wings. More precisely, the TBA equations divided by the asymptotic solution take
exactly the same form. The only differences are in the asymptotic solution and a projection\footnote{Namely, $Y_{a,s}(u)=Y_{a,-s}(-u)$.}. That projection is irrelevant for the vacuum in which all Y-functions are even.

One important feature of the momentum carrying Y-functions is that they have a double pole at zero mirror momentum for any value of the coupling, associated with the exchange of a single particle \cite{Ghoshal:1993tm}.  In the near BPS limit all the dynamics of the momentum carrying Y-functions reduces to the residue of that pole ${\mathbb C}_a$. That is
\beq\la{defC}
\lim_{\substack{u\to0}}\(u^2 Y_{a,0}\)
={1\over4}\phi^4\,{\mathbb C}_{a}^2
+{\cal O}(\phi^6)
\eeq
where $\phi^2$ is considered to be the smallest parameter in the problem. As a result, the cusp TBA system simplifies. Only the leading orders in the expansions of the other Y-functions are relevant. These are
denoted by
\beq\la{expansion}
Y_{1,1}\equiv -1-\phi^2\,\Psi\ \quad\frac{1}{Y_{2,2}}\equiv -1-{\phi^2}\,\Phi\ ,\quad Y_{m,1}{\Big|}_{\phi=0}\!\!={1\over Y_{1,m}}{\Big|}_{\phi=0}\!\!\equiv\cY_m\ ,\quad\frac{1}{Y_{1,m}Y_{m,1}}-1\equiv \phi^2\;{\cal X}_m
\eeq
The corresponding reduced TBA equations become \cite{Correa:2012hh}
\beqa
\Phi-\Psi&=&\sum_{a=1}^\infty\pi\widehat K_a\,{\mathbb C}_a\la{TBA1}\\
\Phi+\Psi&=&{\frak s}*\[-2\frac{{\cal X}_2}{1+{\cal Y}_2}+\sum_{a=1}^\infty\pi (\widehat K^+_a-\widehat K^-_a) \,{\mathbb C}_a-\pi\delta(u)\,{\mathbb C}_1\]\la{TBA2}\\
\la{Ym}\log{{\mathbb Y_m}}&=&I_{m,n}\,{\frak s}*\log\(\frac{\mathbb Y_n}{{1+\mathbb Y_n}}\)
+\delta_{m,2}\, {\frak s}\hat *\[\log\frac{\Phi}{\Psi}+\phi^2(\Phi-\Psi)\]
+\phi^2\pi\, {\frak s}(u) {\mathbb C}_m\la{TBA3}\\
{\mathbb C}_a&=&(-1)^aa^2\,F_a\(\sqrt{1+{a^2\over16g^2}}-{a\over4g}\)^{2+2L}e^{\Delta_a}\la{TBA4}\\
\Delta_a&=&\left.\[\frac{1}{2} K_a\hat* \log\frac{\Psi}{\Phi} +\frac{1}{2}\widetilde K_a\hat * \log(\Psi\Phi)
+\sum_{b=2}^\infty\widetilde K_{ab}*\log\(1+{\cal Y}_b\)-\log a\]\right|_{u=0}\la{TBA5}
\eeqa
where\footnote{Here, $F_a$ is obtained from the one in \cite{Correa:2012hh} by changing the integration variable to $u$.}
\beq\la{bbYandF}
{\mathbb Y_s}\equiv {{\cal Y}_s}(1+\phi^2{{\cal X}_s})\qquad\text{and}\qquad \log F_a=\left.\widetilde K_a\hat *\log
\frac{\sinh(2\pi u)}{2\pi u}\right|_{u=0}
\eeq
The hat on $\hat *$ is a convolution over the range $-2g<u<2g$. Our conventions for the kernels are
\beq\la{thekernels}
\!K_a(u)={2a\over\pi(a^2+4u^2)}\ ,\quad\widehat K_a(u)=\sqrt{4g^2-u^2\over4g^2+a^2/4}K_a(u)\ ,\quad\widetilde K_a(u)=\sqrt{4g^2+a^2/4\over4g^2-u^2}K_a(u)
\eeq
and $\widetilde K_{ab}$ is given in appendix \ref{kernelsconventions}. The cuts in $\widetilde K_a$ and $\widehat K_a$ are chosen to be between $-2g$ and $2g$.
These integral equations are supplemented by boundary conditions. That is, at large real $u$ the function ${\mathbb Y}_m$ approaches its asymptotic value
\beq\la{boundaryconditions}
{\mathbb Y}_m\to{1\over m^2-1}\(1-\phi^2{m^2\over3}\)\;.
\eeq

Finally, the energy is also dominated by the value of $Y_{a,0}$ at the double pole and reduces to
\beq\label{EfromC}
{\cal E}=-\phi^2 B(\lambda)\ ,\qquad B(\lambda)=-\frac{1}{2}\sum_{a=1}^\infty{\mC_a\over\sqrt{1+{16g^2/ a^2}}}\,.
\eeq

Note that even though the reduced TBA equations (\ref{TBA1})-(\ref{TBA5}) are simpler than the general ones \cite{Correa:2012hh}, still they are an infinite set of non-linear integral equations. Moreover, for $L=0$ the ${\mathbb C}_a$'s become oscillatory and truncating their sums becomes problematic for numerics. In the next section we will reduce that infinite set of equations to a finite set (FiNLIE) along the lines of \cite{Gromov:2011cx}. As we will see, doing numerics with that finite set of equations is very simple, even for $L=0$.

\section{Reduction to FiNLIE}
\la{sec:red}

We will now reduce the Bremsstrahlung TBA equations (\ref{TBA1}) -- (\ref{TBA5}) into a few functional equations for only three functions.
We will mainly follow the methods developed in \cite{talk,Gromov:2011cx}, based on Y-system or equivalently on the underlying integrable Hirota dynamics. We will start by applying the method to the infinite set of equations for ${\mathbb Y}_m$ (\ref{TBA3}).

\subsection{Reduction of Infinite Set of Equations for ${\cal Y}_m$}

We first consider the equation for the leading order of ${\mathbb Y}_{m}$, which are denoted as ${\cal Y}_s$ (\ref{bbYandF}). The integral equation for them is
\beq\la{larho}
\log{{{\cal Y}_m}}=I_{m,n}\,{\frak s}*\log\(\frac{{\cal Y}_n}{{1+{\cal Y}_n}}\)
+\delta_{m,2}\, {\frak s}\hat *\log\frac{\Phi}{\Psi}\;.
\eeq
This is exactly the type of equation considered in \cite{Gromov:2008gj,talk,Gromov:2011cx}. That is, the integral equation (\ref{larho}) is equivalent to the functional equation
\beq\la{functional}
\log\(\cY_m^+\cY_m^-\)=I_{m,n}\log{\cY_n\over1+\cY_n}+\delta_{m,2}\log{\Phi\over\Psi}
\eeq
plus the assumption that $\cY_m$ is analytic inside the strip $|{\rm Im}\,u|\le1/2$. This is the Y-system for one dimensional wing (with the source $\log\Phi/\Psi$ at its boundary). This equation can be solved in two steps.
First, one introduces T-functions
\beq\la{YTmap}
{\cal Y}_m=\frac{{\cal T}_{m}^+{\cal T}_{m}^-}{{\cal T}_{m+1}{\cal T}_{m-1}}-1\;.
\eeq
The Y-system equation for the Y-functions is equivalent to the Hirota equation for the T-functions. The advantage of using the T-functions is that the general solution to the Hirota equation is known (see \cite{Gromov:2011cx} for more details). That is, the functional equation (\ref{functional}) is automatically satisfied for $m>2$ provided that ${\cal T}_{s}$ can be written in the form
\beq\la{TtoQ}
{\cal T}_{s}=
\left|
\bea{cc}
Q_1(u+\tfrac{is}{2}) & Q_2(u+\tfrac{is}{2}) \\
Q_{\bar 2}(u-\tfrac{is}{2}) & Q_{\bar 1}(u-\tfrac{is}{2})
\eea
\right|
\eeq
for any four functions $Q_1,Q_2,Q_{\bar 2},Q_{\bar 1}$. This change of variables thus already reduces the infinite tower of the ${\cal Y}$-functions to the set of four Q-functions.

To reduce further the set of independent functions one can notice that the map \eq{YTmap}
is not one--to--one. Indeed there is a ``gauge" freedom in the choice of the ${\cal T}_{s}$
for a given ${\cal Y}_{s}$. This freedom can be used to set two of the Q-functions to $1$
\beq
Q_2=Q_{\bar 1}=1
\eeq
Furthermore, due to reality of the Y-functions, we can always choose the remaining functions to be complex conjugate of each other
\beq\la{TQrelation}
 Q_{1}=-\bar Q_{\bar 2}\equiv Q\qquad\text{and therefore}\qquad{\cal T}_m=Q(u+\tfrac{im}{2})+\bar Q(u-\tfrac{im}{2})\ .
\eeq
This leaves us with essentially one single function!

Next, we have to ensure the analytical properties that follows from the integral TBA equations (\ref{larho}). One of these is the analyticity of ${\cal Y}_m$ inside the strip $-\frac{m-1}{2}<\IM u<\frac{m-1}{2}$. This property implies that
one can always choose the remaining function $Q$ to be analytic everywhere above the real axis. As a result, we can write the following spectral representation for it
\beq
Q(u)=-i u -\int\limits_{-\infty}^\infty\!\!\frac{dv}{2\pi i}\frac{\rho(v)}{u-v}\qquad\text{for}\qquad \IM u>0\;.
\eeq
Here, the first term, $-iu$, follows from the asymptotic behavior  of the Y-functions at large real $u$. The overall normalization is of course a gauge choice.

Another analyticity constraint follows from the TBA equation for $m=2$. It follows from that equation that ${\cal Y}_m$ should have at least two branch cuts starting at $u=\pm(i/2+2g)$
\cite{Cavaglia:2010nm,Gromov:2011cx}.
This is automatically true if $\rho(v)$ is nonzero inside the interval $[-2g,2g]$ and vanishes outside. That is, the ends of the integration interval will create
branch points in $Q$ at $u=\pm 2g$. These cuts then translate through (\ref{TtoQ}) into cuts at $u=\pm(i/2+2g)$ for ${\cal T}_1$ and hence for $\cY_2$ (\ref{YTmap}).
In what follows, we will assume that the function $\rho$ is real. As we will see, that assumption is consistent with the equations (see also \cite{Gromov:2011cx} for a proof related to ${\mathbb Z}_4$ symmetry of the worldsheet action).

To summarize, the infinite set of integral equations (\ref{larho}) can be packed into one single finite support function $\rho(u)$ as
\beq\la{ansatz}
{\cal T}_m=m+K_m\hat *\rho\;.
\eeq
The function $\rho$ is fixed from the functional equation (\ref{functional}) at  $m=2$ and the boundary conditions (\ref{boundaryconditions}). In terms of $\rho$, that equation reads
\beq\la{PoP}
\boxed{\frac{\Phi}{\Psi}=
\frac{{\cal T}_1(u+i/2+i0){\cal T}_1(u-i/2-i0)}{{\cal T}_1(u+i/2-i0){\cal T}_1(u-i/2+i0)}
=
\frac{(1+\sK_1^+\hat*\rho-\frac{1}{2}\rho)(
1+\sK_1^-\hat *\rho-\frac{1}{2}\rho)}{
(1+\sK_1^+\hat*\rho+\frac{1}{2}\rho)(
1+\sK_1^-\hat *\rho+\frac{1}{2}\rho)}}\;.
\eeq
Here $\sK_1^\pm$ denotes the principal value integration with the kernel $K_s(u\pm i/2)$.
The derivation of this equation is straight forward and is described in detail in \cite{Gromov:2011cx}.

\subsection{Reduction of Infinite Set of Equations for ${\cal X}_m$}

Next, we move to the order $\phi^2$ piece of ${\mathbb Y}_m$, namely ${\cal X}_m$ (\ref{bbYandF}). At order $\phi^2$ there are two new sources in the right hand side of (\ref{TBA3})
\beq\la{Xsource}
\phi^2\delta_{m,2} \;{\frak s}\hat *\[\Phi-\Psi\]\qquad\text{and}\qquad\phi^2\pi {\frak s}(u) {\mathbb C}_m\;.
\eeq
The first term is of the same form as the source in the equation for ${\cal Y}_m$ (\ref{larho}). It shifts the left hand side of (\ref{PoP}) by $\phi^2[\Phi-\Psi]$ and therefore is accounted for by a corresponding
shift of the density $\rho\to\rho+\phi^2\varrho$.

To relate the integral equation for $\cY_m$ (\ref{larho}) to the functional Y-system equation (\ref{functional}), we assumed that $\cY_m$ is analytic inside the strip $|{\rm Im}\,u|\le1/2$. That is, for an analytic function $f(u)$ inside the strip (including boundaries) we have
\beq\la{fpms}
f={\frak s}*g\quad\Longleftrightarrow\quad f^++f^-=g\;.
\eeq
Suppose we modified the function $f(u)\to\tilde f(u)$ so that it now has poles at $u=\pm i/2$
\beq
\tilde f(u)\equiv f(u)+\frac{i{\mathbb C}/2}{u+{i\over2}}-\frac{i{\mathbb C}/2}{u-{i\over2}}= f(u)+{\mathbb C}\pi K_1(u)\;.
\eeq
Due to the contributions of these poles at the boundary of the strip, (\ref{fpms}) is modified to
\beq
\tilde f={\frak s}*\tilde g+\pi\,{\mathbb C}\,{\frak s}(u)
\eeq
where
\beq
\tilde g(u)\equiv\tilde f^+(u-i0)+\tilde f^-(u+i0)=g+\pi\,{\mathbb C}K_2(u)+\pi\,{\mathbb C}\,\delta(u)
\eeq
The equation for ${\mathbb Y}_m$, (\ref{TBA3}) is exactly of that form! That is, the second term in (\ref{Xsource}) tells us that ${\mathbb Y}_m$ should have poles at $u=\pm i/2$ with residues $\mp i\phi^2{\mathbb C}_m/2$ correspondingly.

Having understood this, our task becomes very simple -- we have to modify our ansatz for ${\cal T}_m$ \eq{ansatz}
to incorporate these poles. That can be easily archived by adding poles at $u={i\over2}\mathbb Z$ to the Q-function
with  to--be--fixed residues
\beq
{\mathbb Q}(u)=-i\[u+\phi^2 {u^3\over3}\] -\int\limits_{-\infty}^\infty\!\!\frac{dv}{2\pi i}\frac{\rho(v)+\phi^2\varrho}{u-v}+{1\over4\pi}\sum_{a\ne0}{b_a\over u-i{a\over2}}\qquad\text{for}\qquad {\rm Im}\,u>0
\eeq
This procedure gives
\beqa\la{dansatz}
{\mathbb T}_m={\cal T}_m+\phi^2 \tau_m\ ,\qquad
\tau_m=\[mu^2-m^3/12\]+K_m\hat*\varrho+\sum_{n=1}^\infty \[b_nK_{n-m}(u)+b_{-n}K_{m+n}(u)\]
\eeqa
where
\beq\la{TtoY}
{\mathbb Y}_m=\frac{{\mathbb T}_{m}^+{\mathbb T}_{m}^-}{{\mathbb T}_{m+1}{\mathbb T}_{m-1}}-1
\eeq
automatically solves the functional Y-system equation. Here, the $b_n$'s are related to the residue of the poles.
The polynomial $-i[u+\phi^2u^3/3]$ in $Q$ is needed to support the boundary conditions at large $u$ (\ref{boundaryconditions}).

It remains to fix the function $\varrho(u)$ and the residues $b_n$.
We fix the residues in the next section.
The equation for $\varrho$ can be obtained by correcting the r.h.s. and l.h.s. in \eq{PoP}
and expanding to leading order in $\phi^2$. It reads
\beq\la{PoP2}
\boxed{\Phi-\Psi=\frac{\tau_1^+\rho-(1+\sK_1^+\hat*\rho)\varrho}{(1+\sK_1^+\hat*\rho)^2-\tfrac{1}{4}\rho^2}+
\frac{\tau_1^-\rho-(1+\sK_1^-\hat*\rho)\varrho}{(1+\sK_1^-\hat*\rho)^2-\tfrac{1}{4}\rho^2}}\;.
\eeq
Note that this equation also contains terms linear in $\tau_1$. It is therefore a linear equation for $\varrho$.

\subsubsection{Fixing Residues of $Q$}

We will now fix the coefficients $b_a$. Following the construction above, $\log{\mathbb Y}_m$ should be analytic in the strip $|{\rm Im}\,u|<1/2$ with only two poles at its boundary $u=\pm i/2$. We have\footnote{All equations above should be understood to linear order in $\phi^2$ only.}
\beq\la{mathbbY}
{\mathbb Y}_m= {\cal Y}_m+\phi^2\,\cY_m\cX_m= {\cal Y}_m+\phi^2 (1+{\cal Y}_m)\[
\frac{\tau_{m}^+}{{\cal T}_{m}^+}
+\frac{\tau_{m}^-}{{\cal T}_{m}^-}
-\frac{\tau_{m+1}}{{\cal T}_{m+1}}
-\frac{\tau_{m-1}}{{\cal T}_{m-1}}
\]\;.
\eeq
The terms containing $\tau_m^+$ and $\tau_m^-$ have potential poles at the origin. The condition that the
residues cancel reads
\beq
\frac{b_{m-1}-b_{m+1}}{{\cal T}_m(i/2)}-\frac{b_{m-1}-b_{m+1}}{{\cal T}_m(-i/2)}=0\;.
\eeq
That condition is automatically satisfied provided that the function ${\cal T}_m$ is even (as the vacuum Y-functions are).

Next, we proceed to the poles at $u=\pm i/2$. The residues at these poles are fixed by \eq{Ym} to be
\beq\la{residue}
\log {\mathbb Y}_m(u)\simeq {{\mathbb C}_m\over2i}\[\frac{\phi^2}{u-{i\over2}}-\frac{\phi^2}{u+{i\over2}}\]\;.
\eeq
This has to be matched with the residue imposed by our ansatz.
Close to $u=i/2$, ${\mathbb Y}_m$ (\ref{mathbbY}) diverges as
\beq\la{bbYexp}
{\mathbb Y}_m(u)\simeq \[1+{ {\cal Y}_m(\tfrac{i}{2})}\] \[
\frac{b_{m+2}-b_m}{{\cal T}_{m+1}(\tfrac{i}{2})}+
\frac{b_m-b_{m-2}}{{\cal T}_{m-1}(\tfrac{i}{2})}-
\frac{b_{m+2}-b_{m-2}}{{\cal T}_{m}(i)}
\]\frac{1}{2\pi i}\frac{\phi^2}{u-\tfrac{i}{2}}\;.
\eeq
By comparing (\ref{bbYexp}) with \eq{residue} we get a recurrence equation for the $b_m$'s
\beq\la{recurrence}
\pi\, {\mathbb C}_m =
\[1+{1 \over { {\cal Y}_m(\tfrac{i}{2})}}\] \[
\frac{b_{m+2}-b_m}{{\cal T}_{m+1}(\tfrac{i}{2})}+
\frac{b_m-b_{m-2}}{{\cal T}_{m-1}(\tfrac{i}{2})}-
\frac{b_{m+2}-b_{m-2}}{{\cal T}_{m}(i)}
\]\;.
\eeq
This equation can be simplified a bit by introducing the numbers
\beq\la{cdefinition}
c_0\equiv\rho(0)\ ,\qquad c_{m\ne0}\equiv \cT_m(0)\;.
\eeq
As $\rho(u)$ is also an even function, the $c_m$'s are real. In terms of these we have
\beq\la{Ytoc}
{\cal T}_m(ia/2)={c_{a+m}-c_{a-m}\over2}\quad\Rightarrow\quad{\cal Y}_m(ia/2)=\frac{(c_{a-m-1}-c_{a-m+1})
   (c_{a+m-1}-c_{a+m+1})}{(c_{a-m+1}-c_{a+m-1})(c_{a-m-1}-c_{a+m+1})}\;.
\eeq
In term of the $c_m$'s, the recurrence relations (\ref{recurrence}) reads
\beq\la{Cb}
\boxed{\frac{\pi{\mathbb C}_m}{c_m}=4\frac{
   b_m -b_{m-2}}{c_m^2-c_{m-2}^2}-4\frac{b_m-b_{m+2}}{c_m^2-c_{m+2}^2}}\ ,\qquad\text{where}\quad b_0=0\quad\text{and}\quad m=2,3,\dots\;.\\
\eeq
This concludes our reduction of the infinite set of integral equations (\ref{TBA3}) for ${\mathbb Y}_m(u)$ to a finite set of equations (\ref{PoP}), (\ref{PoP2}), (\ref{Cb}) for $\rho(u)$, $\varrho(u)$ and for the coefficients $b_m$. In the next section we will rewrite the
remaining TBA equations by plugging our ansatz (\ref{dansatz}) into them.

\subsection{Simplified Equation for $\Delta_m$}
In this section we will analyze the equation for $\Delta_m$ (\ref{TBA5}). Using the ansatz of the previous section for ${\mathbb Y}_m$, we will get rid of the infinite sums in (\ref{TBA5}). That procedure is called ``telescoping" in \cite{Gromov:2011cx}.

One of the terms in the equation for $\Delta_m$ is $\widetilde K_{ab}*\log\(1+\cY_b\)$.
Using \eq{YTmap}, those terms read
\beqa
&&\!\!\!\!\!\sum_{b=2}^\infty\widetilde K_{ab}*\log(1+{\cal Y}_b) =
\sum_{b=2}^\infty\widetilde K_{ab}*\[\log{{\cal T}_b^+}{}
+
\log{{\cal T}_b^-}{}
-
\log{{\cal T}_{b+1}}{}
-
\log{{\cal T}_{b-1}}{}\]\\
&&\!\!\!\!\!=
\sum_{b=2}^\infty\[
\widetilde K_{ab}^{+_-}+\widetilde K_{ab}^{-_+}-
\widetilde K_{a,b-1}
-
\widetilde K_{a,b+1}\]*
\log{{\cal T}_{b}}+\widetilde K_{a,1}*
\log{{\cal T}_{2}}{}
-\widetilde K_{a,2}*
\log{{\cal T}_{1}}{}\nn
\eeqa
where we introduce a new notation for the shift $f^{\pm_{\Blue{\pm}}}\equiv f(u\pm i/2\Blue{\pm i0} )$
so that in the first two terms the shifts are $\pm i/2\mp i0$ . The last two terms appear when we shift the summation index by $1$.
In fact $\widetilde K_{a,1}=0$, so only one of these is non-zero.

Now we notice something nice -- the expression in the square brackets is almost zero.
It is nonzero only due to the small displacements by $\pm i0$ which may produce a delta function.
Furthermore, one should be careful with the boundary case $b=2$ where the branch cuts in $\widetilde K_{a,2}^\pm$ approach the real
axis and we get an extra contribution due to the square root cut
\beq
\[\widetilde K_{ab}^{+_-}(u)
+
\widetilde K_{ab}^{-_+}(u)
-
\widetilde K_{a,b-1}(u)
-
\widetilde K_{a,b+1}(u)\]
=
\delta_{a,b}\,\delta(u)+\delta_{b,2}\,\chi(u)\,\widetilde K_a(u)
\eeq
where $\chi(u)\equiv 1$ for $-2g<u<2g$ and is zero otherwise. Thus the above integral simply gives
\beq
\sum_{b=2}^\infty\widetilde K_{ab}*\log(1+{\cal Y}_b)=
\log{\cal T}_{a}-
\widetilde K_{a,2}*
\log{{\cal T}_{1}}
+\widetilde K_a\hat *\log{\cal T}_{2}\;.
\eeq
Let us rewrite the term $\widetilde K_{a,2}*
\log{{\cal T}_{1}}$ by writing the kernel more explicitly
and shifting the integration contour
\beqa\la{Ka2logT1}
-\widetilde K_{a,2}*
\log{\cal T}_{1}
&=&
-\frac{1}{2}\(\widetilde K_a^+-\widetilde K_a^-+K_a^++K_a^-\)*
\log{\cal T}_{1}\\
&=&
-\frac{1}{2}\int\limits_{-\infty\Blue{+i0}}^{\infty\Blue{+i0}}\!\!\!\!du\(\widetilde K_a+K_a\)\log {\cal T}_1^{-}
+\frac{1}{2}\int\limits_{-\infty\Blue{-i0}}^{\infty\Blue{-i0}}\!\!\!\!du\(\widetilde K_a-K_a\)\log {\cal T}_1^{+}\nn
\eeqa
Note that if the sign of the $\Blue{i0}$'s was opposite, (\ref{Ka2logT1}) was just zero. That is
\beq
\int\limits_{-\infty\Blue{-i0}}^{\infty\Blue{ -i0}}\!\!\!\!du\(\widetilde K_a+K_a\)\log {\cal T}_1^{-}=0\qquad\text{and}\qquad
\int\limits_{-\infty\Blue{+i0}}^{\infty\Blue{+i0}}\!\!\!\!du\(\widetilde K_a-K_a\)\log {\cal T}_1^{+}=0
\eeq
because the integration contour can be deformed to
infinity as the integrant is analytical
below/above the real axis. The difference between the two signs of $\Blue{i0}$ is the contribution of the square root cut. That is, we have
\beq\la{tKa2T1}
-\widetilde K_{a,2}*
\log{\cal T}_{1}=
\frac{1}{2}K_a\hat *\log \frac{{\cal T}_1^{-_\Blue{-}}{\cal T}_1^{+_\Blue{+}}}{{\cal T}_1^{-_\Blue{+}}{\cal T}_1^{+_\Blue{-}}}-
\frac{1}{2}\widetilde K_a\hat *\log\({{\cal T}^{-_\Blue{+}}_1{\cal T}_1^{+_\Blue{-}}{\cal T}_1^{-_\Blue{-}}{\cal T}_1^{+_\Blue{+}}}\)
\eeq
where again
\beq
{\cal T}^{+_\Blue{\pm}}(u)=\cT(u+i/2\,\Blue{\pm\, i0})\qquad\text{and}\qquad \cT^{-_\Blue{\pm}}(u)=\cT(u-i/2\,\Blue{\pm\, i0})\;.
\eeq
In writing (\ref{tKa2T1}), we used the facts that the square root cut is taken between $-2g$ and $2g$. That is,
\beq
\widetilde K_a(u+i0)=\left\{\begin{array}{llc}-\widetilde K_a(u-i0)&\text{for}&-2g<u<2g\\
+\widetilde K_a(u-i0)&\text{for}& u<-2g\;\|\;2g<u\end{array}\right.\ .\nn
\eeq

Now, the first term on the r.h.s. of (\ref{tKa2T1}) is nothing but $\tfrac{1}{2}K_a\hat *\Phi/\Psi$ according to \eq{PoP}. It cancels precisely with the $\Phi/\Psi$ term in \eq{TBA5}! That is (\ref{TBA5}) reduces to
\beq\la{Delt2}
{\Delta}_a=
\left.\frac{1}{2}\widetilde K_a\hat * \log\frac{\Psi\Phi\,\cT^2_2}{{\cal T}^{-_+}_1{\cal T}_1^{+_-}{\cal T}_1^{-_-}{\cal T}_1^{+_+}}
+\log\frac{{\cal T}_{a}}{a}\right|_{u=0}\;.
\eeq
We will see later that it is highly advantageous to introduce a new notation
\beq\la{et1}
\eta \equiv
\frac{\Psi\,{\cal T}_2}{{\cal T}^{-_+}_1\,{\cal T}_1^{+_-}}=\frac{\Phi\,{\cal T}_2}{{\cal T}^{-_-}_1\,{\cal T}_1^{+_+}}
\eeq
where the second equality in nothing but \eq{PoP}.
Furthermore, utilizing that ${\cal T}_{a}(0)=c_a$ we have
\beqa\la{Delt3}
\boxed{{\Delta}_a=
\widetilde K_a\hat * \log\eta+\log \frac{c_a}{a}}\;.
\eeqa
That is a much nicer equation with no infinite sums!

\subsection{Equations for Fermions ($\Psi$ and $\Phi$)}

Our goal is to have a closed set of equations for a minimal set of functions. We will now exchange the two ``fermionic" functions\footnote{The name is purely historical and does not refer to
Grassman variables.} $\Psi$ and $\Phi$ for the single function $\eta$ (\ref{et1}). As we discussed in the previous section there are two equivalent definitions of $\eta$ (due to \eq{PoP})
\beq\la{et3}
\eta =
\Psi\frac{{\cal T}^{\{-\}}_1+{\cal T}^{\{+\}}_1}{({\cal T}^{\{-\}}_1+\frac{1}{2}\rho)({\cal T}^{\{+\}}_1+\frac{1}{2}\rho)}=
\Phi\frac{{\cal T}^{\{-\}}_1+{\cal T}^{\{+\}}_1}{({\cal T}^{\{-\}}_1-\frac{1}{2}\rho)({\cal T}^{\{+\}}_1-\frac{1}{2}\rho)}
\eeq
where we rewrote \eq{et1} in terms of the principal value 
\beq
{\cal T}^{\{\pm\}}_1\equiv{1\over2}\(\cT_1^{\pm_+}+\cT^{\pm_-}\)= 1+\dashint\limits_{-2g}^{2g}\!\! dv\,\rho(v)\, K_1(u-v\pm \tfrac{i}{2})\;.
\eeq
The functions $\Psi$ and $\Phi$ enter into the equations through their sum and difference. We can now use \eq{et3} to express them in terms of $\rho$ and $\eta$
\beq\la{ferequations}
\boxed{\Psi-\Phi=\rho\,\eta\ ,\qquad
\Psi+\Phi=\eta\frac{\frac{1}{2}\rho^2+2{\cal T}^{\{-\}}_1{\cal T}^{\{+\}}_1}{{\cal T}^{\{-\}}_1+{\cal T}^{\{+\}}_1}}\;.
\eeq
What we have achieved by that is a set of closed equations for only three functions $\{\rho,\eta,\varrho\}$ and one infinite set of real numbers $\{{\mathbb C}_a\}$, as we will now summarize.

\subsection{Summary of Bremsstrahlung FiNLIE}
In this section we summarize our finite set of nonlinear integral equations (FiNLIE) for the functions $\eta$, $\rho$ and $\varrho$. As we will see, only the value of these functions on the interval $[-2g,2g]$ will matter. This is particularly neat for the numerical implementation,
which we discuss in the next section.

The finite set of non-linear equations for $\eta$, $\rho$ and $\varrho$ reads
\vspace{.3cm}
\newcounter{storedequation}
\setcounter{storedequation}{\arabic{equation}}
\setcounter{equation}{0}
\renewcommand{\theequation}{{\textbf{F\arabic{equation}}}}
{}\\
\noindent
\fbox{
\addtolength{\linewidth}{-2\fboxsep}
\addtolength{\linewidth}{-2\fboxrule}
\begin{minipage}{\linewidth}
{}\vspace{-3mm}
\beqa
\la{F1}\rho\!\!\!&=&\!\!\!-\frac{1}{\eta}\sum_{a=1}^\infty \pi\, {\mathbb C}_a\,\widehat K_a\\
\la{F2}\eta\!\!\!&=&\!\!\!
\frac{{\cal T}^{\{-\}}_1+{\cal T}^{\{+\}}_1}{\frac{1}{2}\rho^2+2{\cal T}^{\{-\}}_1{\cal T}^{\{+\}}_1}
\,\times\,{\frak s}*
\[-2\frac{{\cal X}_2}{1+{\cal Y}_2}+\pi (\widehat K^+_a-\widehat K^-_a) \,{\mathbb C}_a-\pi\delta(u)\,{\mathbb C}_1\]\\
\la{F3}\varrho\!\!\!&=&\!\!\!\rho\frac{\tau_1^{\{+\}}\(\frac{\rho ^2}{4}-{\T_1^{\{-\}}}^2\)+\tau_1^{\{-\}}\(\frac{\rho ^2}{4}-{\T_1^{\{+\}}}^2\)-\frac{\eta}{4} \(\frac{\rho ^2}4-\T_1^{\{-\}\;2}\)\(\frac{\rho ^2}4-{\T_1^{\{+\}}}^2\)}
{\(\T_1^{\{-\}}+\T_1^{\{+\}}\)\(\frac{\rho ^2}4-\T_1^{\{-\}}\T_1^{\{+\}}\)}\\
\la{F4}
{\mathbb C}_a\!\!\!&=&\!\!\!(-1)^aa\, c_a\(\sqrt{1+{a^2\over16g^2}}-{a\over4g}\)^{2+2L}\exp\[\widetilde K_a\hat * \log\(\eta
\frac{\sinh(2\pi u)}{2\pi u}\)\]
\eeqa
\vspace{.0cm}
\end{minipage}
}
\vspace{.3cm}\\ 
For a compact summary of our notations we refer the reader to appendix \ref{kernelsconventions}.
Here, (\ref{F1}) and (\ref{F2}) are the fermionic equations \eq{TBA1} and \eq{TBA2} written in terms of $\rho$ and $\eta$.  \setcounter{equation}{\arabic{storedequation}}
\renewcommand{\theequation}{\arabic{equation}}
The combination $\cX_2/(1+\cY_2)$ is given by (\ref{mathbbY})
\beq\la{X2Y2}
\frac{{\cal X}_2}{1+{\cal Y}_2}=-\frac{\T_3\T_1}{\T_1^{++}\T_1^{--}}
\(\frac{\tau_3}{\T_3}+\frac{\tau_1}{\T_1}
-\frac{\tau_2^+}{\T_2^+}-\frac{\tau_2^-}{\T_2^-}\)\;.
\eeq
Equation (\ref{F3}) for $\varrho$ is obtained from \eq{PoP2} using (\ref{ferequations}).
Finally equation (\ref{F4}) for the infinite set of coefficient ${\mathbb C}_a$ is \eq{TBA4} where we used \eq{Delt3} for $\Delta$.
Note that the function $\eta$ has combined neatly with $F_a$ from the boundary dressing phase (\ref{bbYandF}).

These equations allow us to find our three main functions $\{\rho,\varrho,\eta\}$ in terms of $\T_m$ and $\tau_m$. These functions themselves can be computed from $\rho,\varrho$ and ${\mathbb C}_a$ through (\ref{ansatz}) and (\ref{dansatz})
\beq\la{ansatz2}
{\cal T}_m=m+K_m\hat *\rho\ ,\qquad
\tau_m=-\frac{m^3}{12}+mu^2+K_m\hat*\varrho+\sum_{n=-\infty}^\infty b_{n} K_{m-n}
\eeq
where the coefficients $c_m$  and $b_m$ are determined by (\ref{cdefinition}) and (\ref{Cb})
\beq
{c_m={\cal T}_m(0)\ \qquad
{b_{a+2}-b_{a}}{}=(c_{a+2}^2-c_a^2)\times
\left\{
\bea{ll}
\sum\limits_{n=1}^{\infty}\frac{\pi{\mathbb C}_{2n-1}}{4c_{2n-1}}
+\sum\limits_{n=a/2+1}^{\infty}\frac{\pi{\mathbb C}_{2n}}{4c_{2n}}
&
,\;\;a\in 2{\mathbb Z}\\
-\sum\limits_{n=0}^{a/2-1/2}\frac{\pi{\mathbb C}_{2n+1}}{4c_{2n+1}}
&,\;\;a\in 2{\mathbb Z}+1
\eea
\right.}
\eeq
Note that we only need $\tau_1$, $\tau_2$ and $\tau_3$. In order to compute these, it is enough to know the difference $b_{a+2}-b_a$:
\beq
\tau_1=-\frac{1}{12}+u^2+K_1\hat*\varrho-\sum_{n=0}^\infty \(b_{n+2}-b_{n}\) K_{n+1}\ ,\quad
\tau_2=\tau_1^+ + \tau_1^-\ ,\quad
\tau_3=\tau_1^{++} + \tau_1^{} + \tau_1^{--}\;.
\eeq

In the next section we discuss how this system of equations can be implemented numerically
and present our numerical results.

\subsection{Numerical Solution of FiNLIE}
The numerical implementation of the Bremsstrahlung FiNLIE system, summarized in the previous section,
is much simpler than the one for the full spectral FiNLIE \cite{talk,Gromov:2011cx}. There are several reasons for that.
First, there are only three functions ($\rho$, $\varrho$, $\eta$), that have to be updated by iterations.
Second, only their values on the cut $[-2g,2g]$ have to be stored.
For example, to efficiently store these functions at each step of iterations one can use a basis of Legendre polynomials, orthogonal on the cut. This makes it really easy to write a code which iterates these equations.

There is however one new complication that appears in the system. The complication only comes about for $L=0$ and is due to the infinite set of
coefficients ${\mathbb C}_a$. For $L=0$ these coefficients do not go to zero for large $a$. As a result, a naive truncation of the equations at some large $a$
would lead to a wrong result.
To overcome this difficulty we have computed the large $a$ expansion of the ${\mathbb C}_a$'s in terms of $\eta$ analytically, up to order $1/a^{20}$. Having an analytic expression for their expansion at large $a$ allows us to compute the tails of the sums over ${\mathbb C}_a$ analytically. The remaining contribution can be truncated efficiently with essentially arbitrary precision.
For the evaluation of the large $a$ coefficients, we use \eq{F4} which gives the ${\mathbb C}_a$'s in terms of
some simple integrals of $\eta$ and $\rho$.
\begin{figure}[ht]
\begin{center}
\includegraphics[scale=0.8]{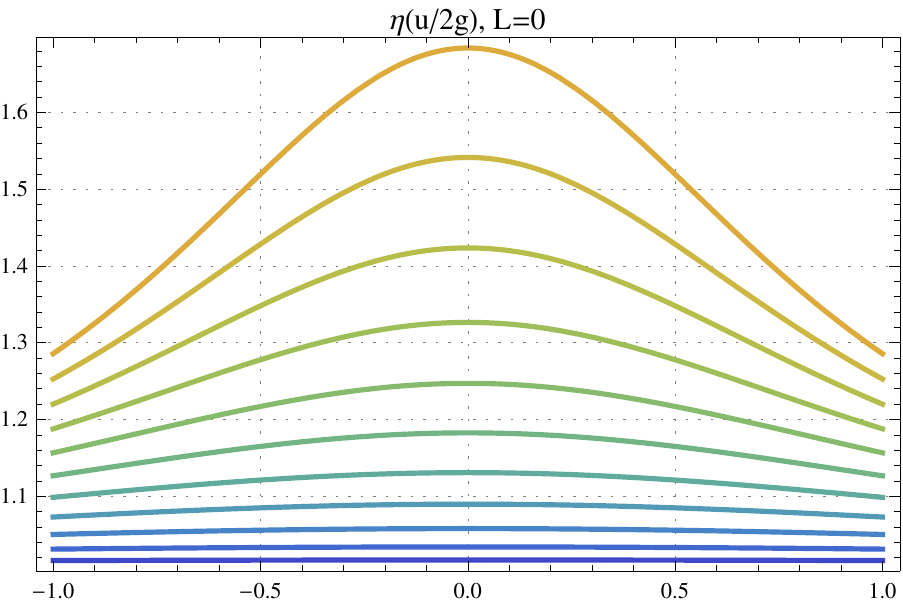}\qquad\qquad
\includegraphics[scale=0.8]{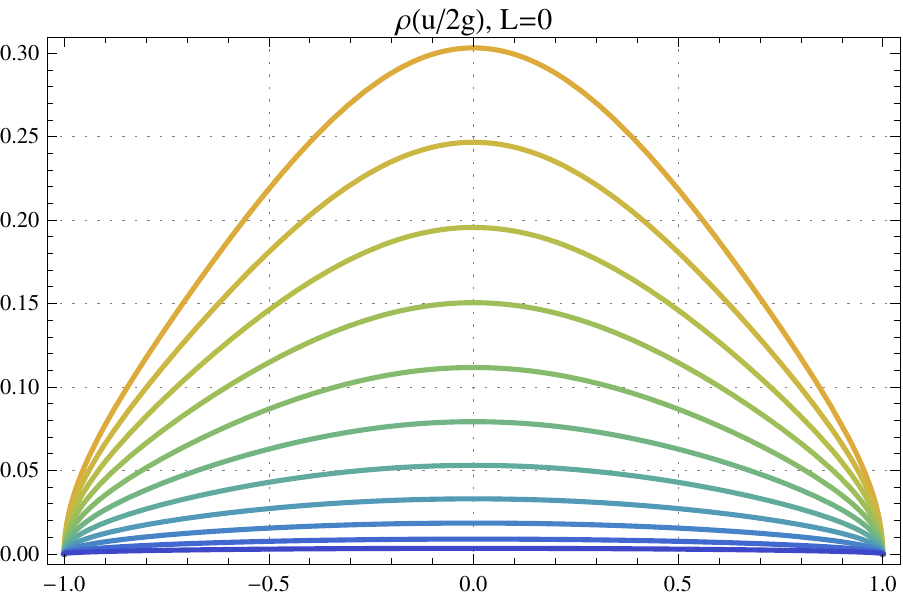}
\end{center}
\caption{\it
Plots of $\eta,\rho$ and $\varrho$ at $L=0$ for various values of coupling from ${g=0.05}$ (bottom/blue) to ${g=0.25}$
(top/red) with a step $\Delta g=0.02$.
\la{figL0}
}
\end{figure}
\begin{figure}[ht]
\begin{center}
\includegraphics[scale=0.8]{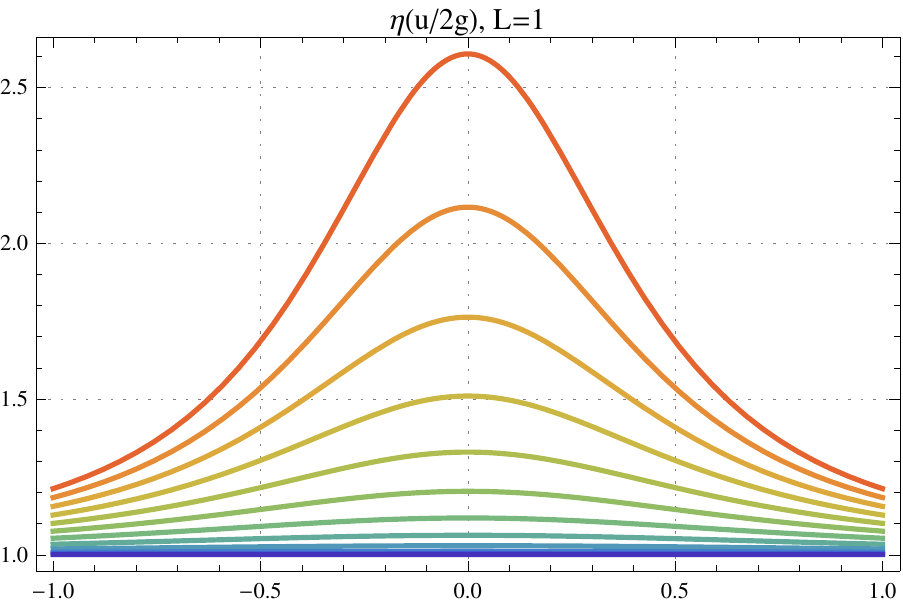}\qquad\qquad
\includegraphics[scale=0.8]{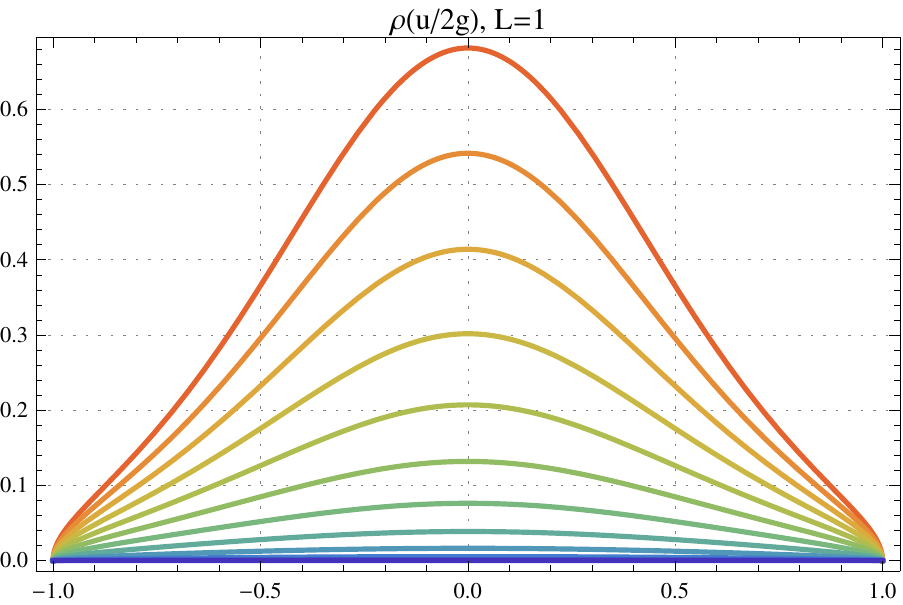}
\end{center}
\caption{\it
Plots of $\eta,\rho$ and $\varrho$ for various values of coupling from $g=0.06$ (bottom/blue) to $g=0.5$
(top/red) with a step $0.04$ with $L=1$.
\la{figL1}
}
\end{figure}

Since solving the TBA equations numerically is a rather standard procedure we describe our
numerical algorithm very briefly. As a base for the iteration process we took the asymptotic values of our functions
$\rho_0=\varrho_0=0$ and $\eta_0=1$.
At the $n$'th iteration step, we first used \eq{F4} to reconstruct several coefficients in the large $a$ expansion of  ${\mathbb C}_a$. Then we plug the current functions
$\rho_n,\;\varrho_n,\;\eta_n$ into the r.h.s. of (\ref{F1}), (\ref{F2}) and (\ref{F3})
to get the next iteration $\rho_{n+1},\;\varrho_{n+1},\;\eta_{n+1}$.
The accuracy of $10^{-7}$ is reached typically after $20$ iterations
and takes about $10$ min on an average modern computer.

\begin{figure}[t]
\begin{center}
\includegraphics[scale=1.2]{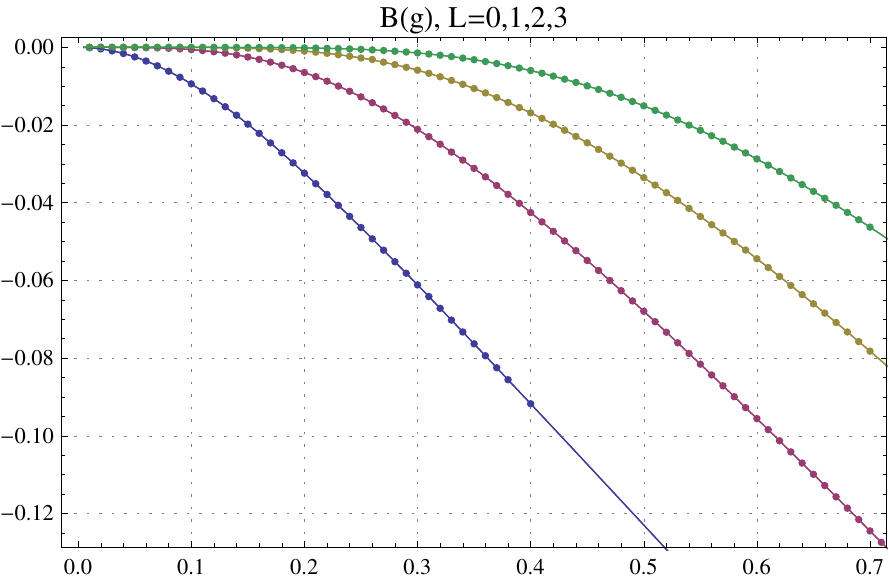}
\end{center}
\caption{\it
Plots of energies for various $L$'s as a function of coupling.
The dots (from bottom to top)
 correspond to $L=0$ (bottom/blue), $L=1$ (red), $L=2$ (yellow), $L=3$ (top/green).
\la{fignum}
}
\end{figure}

The output of the iteration process are the functions $\eta,\rho$
which are plotted in figure~\ref{figL0} for $L=0$ and in figure~\ref{figL1} for $L=1$. The goal of the iteration process is of course the energy. It is given in terms of the ${\mathbb C}_a$'s through (\ref{EfromC}). In figure~\ref{fignum} we presented the result for $L=0,1,2,3$ for a wide range of couplings. In that figure, we also plotted our analytical results (solid lines)
which are derived in the next section.

\section{Analytic Thermodynamic Bethe Ansatz}
\la{sec:ana}

In this section we solve the FiNLIE analytically. We do so by understanding the analytical properties of
our basic quantities $\eta$ and $\rho$.

\subsection{Analytic Ansatz}
In the previous section we formulate a closed system of equations
defined on the interval $u\in[-2g,2g]$. It turns out to be very useful to
think more broadly and analyze the analytical properties of these functions on the whole
complex plane, where the power of complex analysis can come to rescue.

Consider first equation \eq{F1}
\beq\la{re}
\rho\,\eta=-\pi\sum_{a=1}^\infty \widehat K_a {\mathbb C}_a\;.
\eeq
It tells us that the combination $\rho\,\eta$
has only two branch points at $\pm 2g$ and numerous poles outside the real axis
at $i a/2,\;a\in{\mathbb Z}$. What we would like to know is what are the analytical properties of
$\rho$ and $\eta$ separately.

Unfortunately, it seems that there is no simple way to answer this question.
At this point we can only argue at the emotional level that it would be very awkward if both $\eta$ and $\rho$ would have some other branch points that annihilate precisely when we multiply these two functions. Note that $\rho$ must have branch points at $\pm 2g$
as being a discontinuity of $\cT$ (this is also clear from \eq{F2} that $\eta$ does not have a cut on the real axis).
Let us then be optimistic and assume the simplest possibility. Namely, we assume that $\eta$ is a meromorphic function on the whole complex plane.
We justify this assumption {\it a posteriori} by consistency with the other equations and by agreement with numerics.
From (\ref{F2}) it follows that $\eta\to1$ at large $u$. Therefore, more explicitly, we assume that
\beq\la{aansatz}
\eta =\prod_a \frac{u-u_a}{u-v_a}\;
\eeq
for some set of roots $\{u_a\}$ and poles $\{v_a\}$.
In the next section we show that $v_a=i a/2$. These poles of $\eta$ are responsible for the poles on the right hand side of (\ref{re}).

\subsection{Fixing Poles and Residues of $\eta$}

We fix the poles and residues of $\eta$. We start from equation \eq{F2} written in the form
\beq
\eta
\frac{\T_1^{\{-\}}\T_1^{\{+\}}+\frac{\rho^2}{4}}{\T_2/2}=
\;{\frak s}*\[-2\frac{{\cal X}_2}{1+{\cal Y}_2}-\eta^+\rho^+
+\eta^-\rho^--\delta(u)\pi {\mathbb C}_1\]\;.
\eeq
Using (\ref{fpms}), we convert it into the functional equation
\beqa
&&\eta^+\frac{(\T_1-\tfrac{\rho^+}{2})(\T_1^{++}+\tfrac{\rho^+}{2})+\frac{(\rho^+)^2}{4}}{\T_2^+/2}
+
c.c.
=
-2\frac{{\cal X}_2}{1+{\cal Y}_2}-\eta^+\rho^+
+\eta^-\rho^-\;.
\eeqa
By plugging in the explicit expression for $\frac{{\cal X}_2}{1+{\cal Y}_2}$ \eq{X2Y2}, on the l.h.s and using that\\ $\cT_1+\cT_1^{++}=\cT_2^+$ on the r.h.s,  we arrive at\footnote{That relation follows from (\ref{TQrelation}).}
\beqa\la{forpoles}
&&\frac{\eta^+ \T_1^{++}}{\T_2^{+}}
+
\frac{\eta^-\T_1^{--}}{\T_2^{-}}
=
-
\frac{\eta^+\rho^+}{\T_2^{+}}
+
\frac{\eta^-\rho^-}{\T_2^{-}}
+
\frac{\T_3}{\T_1^{++}\T_1^{--}}\(\frac{\tau_3}{\T_3}+\frac{\tau_1}{\T_1}-\frac{\tau_2^+}{\T_2^+}-\frac{\tau_2^-}{\T_2^-}\)\;.
\eeqa

This form is very useful for analyzing the poles of $\eta$. First, note that $\cT_m\simeq m$ and therefore $\T_m$ can not have zeros at sufficiently small $g$.\footnote{The
function $\cT_m$ only depends on $\rho$ in the interval $[-2g,2g]$. Moreover, at $g\to0$ the solution reduces to the asymptotic one (\ref{boundaryconditions}) where
$\eta\simeq 1$, $\rho\simeq 0$ and therefore $\cT_m\simeq m$.} As the function $\eta$ is analytic in $g$, the only potential poles in (\ref{forpoles}) are the following -- poles of $\tau_s$
and $(\eta\rho)^\pm$ on the r.h.s and poles of $\eta^\pm$ on the l.h.s.
Both type of terms on the r.h.s can only have poles at $ia/2$ for some integer $a$. It follows that also $\eta$ has poles at $ia/2$.
However, this does not mean that all the poles of $\eta$ are at $ia/2$ since there could be
another infinite series of poles which cancel in the finite difference. This possibility
is excluded below. At the moment let us write a recursive relation for the residues of $\eta$
at $ia/2$. We denote them by $e_a$
\beq
\underset{u=ia/2}{\rm Res}\;\eta = \frac{e_a}{2\pi i}\;.
\eeq
Using that
\beq
\underset{u=ia/2}{\rm Res}\;\rho\,\eta = -\frac{\pi{\mathbb C}_a}{2\pi i}\ ,\qquad
\underset{u=ia/2}{\rm Res}\;\tau_s = \frac{b_{a-s}-b_{a+s}}{2\pi i}\ ,\qquad
{\cal T}_m(ia/2)={c_{a+m}-c_{a-m}\over2}
\eeq
we get
\beq\la{polefix}
\frac{\left(c_{a-1}-c_{a-3}\right) \left(c_{a-1} e_{a-1}+{\pi\,\mathbb C}_{a-1}\right)}{c_{a-1}
   \left(c_{a+1}-c_{a-3}\right)}+\frac{\left(c_{a+1}-c_{a+3}\right)
   \left(c_{a+1} e_{a+1}+{\pi\,\mathbb C}_{a+1}\right)}{c_{a+1}
   \left(c_{a-1}-c_{a+3}\right)}=0\;.
\eeq
It is very appealing to solve this second order recurrence equation  by setting $e_a=-\pi\, C_{a}/c_a$. However, we have to exclude
other possible solutions by fixing the initial conditions.

Consider first (\ref{polefix}) at $a=1$. We have ${\mathbb C}_0\equiv 0$ as the sum in (\ref{re}) starts from $a=1$. Moreover, $e_0=0$. Otherwise, it would follow from (\ref{re}) that $\rho(0)=0$, which generically is not true. Hence, the only solution is $e_2=-\frac{\pi C_2}{c_2}$. As a consequence, for any {\it even} $a$ we have $e_a=-\frac{\pi C_a}{c_a}$.

It is left to fix the residues $\{e_a\}$ at odd $a$'s. The first of these, for $a=\pm1$, are at the boundary of the strip $-1<\IM u<1$. Therefore, we go back to the integral equation \eq{F2}. It gives us direct accesses to the analytical
properties of $\eta$ inside this strip. The only source for the poles in the r.h.s. of
\eq{F2} is the term with $\delta$-function (which so far played no role). It gives
\beq
\eta
\frac{\T_1^-\T_1^++\frac{\rho^2}{4}}{\T_2/2}\simeq
-\frac{\pi {\mathbb C}_1}{2\cosh(\pi u)}
\eeq
comparing residues at $i/2$ of right and left hand sides we get
\beq
\frac{c_1 e_1\frac{c_3-c_1}{2}-\frac{\pi C_1}{2}(c_1-\frac{c_3-c_1}{2})}{\frac{c_3+c_1}{4}}=
{-\pi C_1}\qquad\Rightarrow\qquad e_1=-\frac{\pi C_1}{c_1}\;.
\eeq
We conclude that for any $a$
\beq
e_a=-\frac{\pi C_a}{c_a}\;.
\eeq
Note that (\ref{F2}) implies that there are no other poles in the strip (\ref{F2}). It follows from (\ref{forpoles}) that there are no other poles in $\eta$ except $ia/2$.
Furthermore, taking into account that form \eq{F2} $\eta\to 1$ for large $u$, we can write the following representation of $\eta$
\beq\la{etafin}
\boxed{\eta(u)=1-\sum_a\frac{\pi {\mathbb C}_a}{c_a}K_a(u)}\;.
\eeq
This is the key equation which will allow us to solve the system analytically. This equation was successfully tested numerically!
In the next sections we also constrain the zeros of $\eta(u)$, thus writing a closed equation for $\eta(u)$ alone.

\subsection{Effective Baxter Equation for Zeros of $\eta$}
In the previous section we have found that the poles of $\eta(u)$ are at $ia/2$.
Then \eq{aansatz} becomes
\beqa\la{aansatz2}
\eta(u)=\prod_{k=1}^\infty\frac{u^2-u_k^2}{u^2+k^2/4}
\eeqa
where we also use that $\eta$ is an even function, so its zeros and poles should generically come in pairs.

The goal of this section is to find the zeroes $u_k$ in closed terms.
For that we plug the ansatz \eq{aansatz2} into \eq{F4}.
We see that we can evaluate the integral in \eq{F4} explicitly in terms of $u_k$.
Using
\beq
{\sinh(2\pi u)\over 2\pi u}=\prod_{k=1}^\infty\frac{u^2+k^2/4}{k^2/4}
\eeq
the exponent of \eq{F4} becomes
\beq
\sum_{k=1}^\infty\frac{1}{2}\oint\limits_{-2g}^{2g}du\,\widetilde K_a\log\frac{u^2-u_k^2}{k^2/4}
\eeq
where we have written the convolution as a contour integral around the cut $[-2g,2g]$.
This allows us to blow the contour to infinity. There are two type of obstacles on that way --
poles of $\widetilde K_a$ at $u=\pm ia/2$ and logarithmic cuts starting at $u=\pm u_k$. By picking the residues at the poles and integrating the discontinuities along the cuts, we obtain
\beqa
\text{poles}\;\;&:&\;\;\log\frac{-a^2/4-u_k^2}{k^2/4}
=
\lim_{u\to ai/2}\log\frac{\eta(u)\sinh(2\pi u)}{2\pi u}=
\log\[(-1)^{a}\frac{{\mathbb C}_a}{a\, c_a}\]
\\ 
\text{cuts}\;\;&:&\;\;
\sum_{k=1}^\infty\log\frac{x_k^2-\frac{1}{y_a^2}}{x_k^2-y_a^2}
\eeqa
where the contribution of the cuts is written in terms of the Zhukovsky variable $x_k=x(u_k)$ and $y_a=x(ia/2)$.\footnote{The Zhukovsky map is defined as ${u\over g}=x+{1\over x}$ and we consider the solution outside the unit circle.}
Combining both contributions together we get from \eq{F4}
\beq\la{F4noeta}
{\mathbb C}_a=(-1)^a a c_a\(\sqrt{1+\frac{a^2}{16g^2}}-\frac{a}{4g}\)^{2+2L}
\[(-1)^{a}\frac{{\mathbb C}_a}{a\, c_a}\prod_{k=1}^\infty \frac{x_k^2-\frac{1}{y_a^2}}{x_k^2-y_a^2}\]\;.
\eeq
The expression in the large round brackets is just $i/y_a$ and (\ref{F4noeta}) simplifies to
\beq
1=\(\frac{i}{y_a}\)^{2+2L}
\[\prod_{k=1}^\infty \frac{x_k^2-\frac{1}{y_a^2}}{x_k^2-y_a^2}\]\;.
\eeq
This strongly reminds some kind of Bethe ansatz equation!
To solve it we use the technology of Baxter equation. Namely, we construct a function
\beq\la{Tdef}
{\bf T}(x)\equiv
 x^{L+1}{\bf Q}(x)+\frac{(-1)^L}{x^{L+1}}{\bf Q}(1/x)\ ,\qquad{\bf Q}(x)\equiv \prod_{k=1}^\infty\frac{x_k^2-{x^2}}{x_k^2}\;.
\eeq
This function has very simple properties. These are
\beq\la{Tprop}
{\bf T}(x)=(-1)^L{\bf T}(1/x)\ ,\qquad {\bf T}(y_a)=0\;
\eeq
and the absence of singularities, apart from essential singularities at $x=0$ and $x=\infty$.
To analyze the behavior at infinity (and at zero) let us relate the Q-functions with $\eta$:
\beq\la{etaQQ}
{\bf Q}(x){\bf Q}(1/x)=\prod_{k=1}^\infty \frac{u_k^2-u^2}{g^2 x^2_k}
=\eta(u)\,{\sinh(2\pi u)\over2\pi u}\widetilde C
\eeq
where the constant $\widetilde C\equiv \prod_{k=1}^\infty\frac{-k^2/4}{g^2 x_k^2}$.
When $y\to\pm\infty$ we also have $u\to\pm\infty$, thus $\eta(u)\simeq {\bf Q}(1/y)\simeq 1$, see (\ref{aansatz2}) and (\ref{Tdef}). Therefore, we have
\beq
{\bf Q}(y)\simeq \,{\sinh(2\pi u)\over2\pi u}\widetilde C\qquad\text{at}\qquad y\to\pm\infty\;.
\eeq
This allows us to estimate the behavior of ${\bf T}$
\beq
{\bf T}(y)\simeq  y^{L}{\sinh(2\pi u)}\frac{\widetilde C}{2\pi g}\qquad\text{at}\qquad y\to\pm\infty\;.
\eeq
The function which enjoys this asymptotics and satisfies \eq{Tprop} is
\beq\la{eqTlarge}
{\bf T}(y)=
\sinh(2\pi u)P(y)\qquad\text{with}\qquad P(y)\equiv
C_1 y^L+C_2 y^{L-2}+\dots+
\frac{C_{1+L}}{y^{L}}
\eeq
where
$C_{L+2-i}=(-1)^L C_i$. Note that $\widetilde C= 2\pi g\, C_1$ and therefore
\beq\la{etaQQ2}
\boxed{\eta(u_x)=\frac{u\;{\bf Q}(x){\bf Q}(1/x)}{g\, C_1 \sinh(2\pi u_x)}}\;.
\eeq
where $u_x\equiv g\(x+{1\over x}\)$.

Finally, recall that that ${\bf Q}(x)=\[P_L(x)\sinh(2\pi u)/x^{L+1}\]_+$, where the subscript ($+$) stands for the regular part of the Laurent expansion. We can therefore write an integral representation for ${\bf Q}$ in terms of $P_L$ as
\beqa\la{PLtoQ}
\boxed{\la{firstQ}{\bf Q}(y)=-\!\!\oint\limits_{|x|>|y|}\!\!{dx\over 2\pi i} \frac{P_L(x)\sinh(2\pi u_x)}{x^2-y^2}{1\over x^L}}
\eeqa

To fix ${\bf T}(x)$ completely, it remains to find the coefficients $C_i$. That is done in the next section.

\subsection{Solution of Effective Baxter Equation}

For simplicity, we will first demonstrate the procedure at $L=0$. Then, we will move to the general case.

\subsubsection{Q-function for $L=0$}
For $L=0$ the general equation \eq{eqTlarge} becomes
\beq
{\bf T}(y)=C_1\,{\sinh(2\pi u_y)}\;.
\eeq
It is useful to expand ${\bf T}$ in Laurent series around $y=0$. For that we use the well known fact that $\sinh$ is a generating function
for Bessel functions
\beq
{\bf T}(y)=C_1\sum_{r=-\infty}^{\infty}I_{2r+1}\(4\pi g\) y^{2r+1}
\eeq
comparing with \eq{Tdef} we see that the first term contains only negatives powers of $y$ whereas the
second term has only positive powers. That is, we have
\beq
{\bf Q}(y)\equiv \prod_{k=1}^\infty\frac{x_k^2-{y^2}}{x_k^2}
=
C_1\sum_{r=0}^{\infty}I_{2r+1}\(4\pi g\) y^{2r}
=-C_1\!\!\oint\limits_{|x|>|y|}\!\!{dx\over 2\pi i}\frac{e^{2\pi u_z}}{x^2- y^2}\;.
\eeq
The constant $C_1$ is fixed from the condition ${\bf Q}(0)=1$. It reads
\beq
C_1={1 \over I_{1}\(4\pi g\)}\;.
\eeq
We will now generalize the construction of ${\bf Q}$ to any $L$.

\subsubsection{Q-function for General $L$}\la{Qfunctionsec}

For general $L$ we had
\beq\la{TtpP}
{\bf T}(y)=
 y^{L+1}{\bf Q}(y)+\frac{(-1)^L}{y^{L+1}}{\bf Q}(1/y)=\sinh(2\pi u_y)P(y)
\eeq
and therefore ${\bf T}(x)$ also has the zeros of $P(y)$ in addition to $y_a=x(ia/2)$.
The coefficients in the Laurent expansion of $P(y)$, $C_r$ (\ref{eqTlarge}) should be determined so that ${\bf T}(y)$ has a gap in its Laurent expansion from $-L$ to $L$
\beq\la{gap}
{\bf T}(y)=\sinh(2\pi u)\sum_{r=0}^{L}C_{r+1} y^{L-2r}=\dots + \frac{(-1)^L}{y^{L+1}}+\frac{0}{y^{L-1}}+\dots+
0\times y^{L-1}+ y^{L+1}+\dots
\eeq
More explicitly, the coefficients $C_j$ satisfy
\beq
\oint{dx\over2\pi i}{P(x)\sinh(2\pi u_x)\over x^{L+2-2i}}=\sum_{j=1}^{L+1} C_jI_{2(j-i)-1} =(-1)^L\delta_{i,L+1}\qquad\text{for}\qquad i=1,\dots,L+1\;.
\eeq
These are just $L+1$ linear equations for $L+1$ variables, the $C_a$'s. We have $C_a=(-1)^L\[{\M}_L^{-1}\]_{a,L+1}$ where
the $(L+1)\times(L+1)$ matrix $\M_L$ is given by
\beq\la{MLmatrix}
\M_L=
\left(\!\!
\bea{ccccc}
I_{-1}&I_{1}&\dots&I_{2L-3}&I_{2L-1}\\
I_{-3}&I_{-1}&\dots&I_{2L-5}&I_{2L-3}\\
\vdots&\vdots& \ddots & \vdots& \vdots\\
I_{1-2L}&I_{3-2L}&\dots&I_{-1}&I_{1}\\
I_{-1-2L}&I_{1-2L}&\dots&I_{-3}&I_{-1}
\eea\!\!
\right)\;.
\eeq

The function $P_L$ can be now written as
\beq\la{PLex}
\boxed
{
P_{L}(x)=\frac{1}{\det{\M}_L}\left|
\bea{ccccc}
I_{-1}&I_{1}&\!\!\dots\!\!&I_{2L-3}&I_{2L-1}\\
I_{-3}&I_{-1}&\!\!\dots\!\!&I_{2L-5}&I_{2L-3}\\
\vdots&\vdots&\!\! \ddots\!\! & \vdots& \vdots\\
I_{1-2L}&I_{3-2L}&\!\!\dots\!\!&I_{-1}&I_{1}\\
1/x^{L}&1/x^{L-2}&\!\!\dots\!\!&x^{L-2}&x^{L}
\eea
\right|
}\;.
\eeq
For example, for $L=0,1,2$ we get
\beqa
P_0=\frac{1}{I_1}\ ,\qquad P_1=\frac{x-1/x}{I_1-I_3}\ ,\qquad P_2=
\frac{I_1 x^2-(I_1+I_3)+I_1/x^2}{I_1^2-I_1 I_3+I_1 I_5-I_3^2}
.
\eeqa
In terms of $z\equiv \tfrac12\(x+1/x\)$, the function $P_L(x(z))$ is a polynomial of degree $L$ for even $L$. Similarly, for odd $L$ the function $P_L(x(z))/[x(z)-1/x(z)]$ is a polynomial of degree $L-1$. In appendix~\ref{OrthogonalPolynomials}, we show that
these are orthogonal polynomials for some simple measure.

Having explicitly the function $P_L$ and therefore also ${\bf Q}$ through (\ref{PLtoQ}), we can now easily reconstruct $\eta(u)$ itself via \eq{etaQQ2}.
In the next section we use the solution for $\eta(u)$ to compute the energy. In that computation, we will also need a function with one less degree than $P_L$. It is defined as $q_L(y)\equiv \[P(y)\sinh(2\pi u)\]_+-P(y)\[\sinh(2\pi u)\]_+$. It follows that $q_L(1/x)=(-1)^{L-1}q_L(x)$ and at the roots of $P_L$ we have $q(y_i)=y_i^{L+1}\bQ(y_i)$, see appendix \ref{qappendix} for details. We find that
\beqa\la{qfunction}
q_{L}(x)=
\small
\left(\!\!
\bea{c}
x^{1-L}\\
x^{3-L}\\
\vdots\\
x^{L-1}\\
x^{L+1}
\eea
\!\!\right)^T\!\!\cdot
\left(\!\!
\bea{ccccc}
0&I_{1}&\dots&I_{2L-3}&I_{2L-1}\\
0&0&\dots&I_{2L-5}&I_{2L-3}\\
\vdots&\vdots& \ddots & \vdots& \vdots\\
0&0&\dots&0&I_{1}\\
0&0&\dots&0&0
\eea
\!\!\right)\cdot\left(\!\!
\bea{c}
C_{L+1}\\
C_L\\
\vdots\\
C_2\\
C_1
\eea
\!\!\right)\;.\la{qfinal}
\eeqa
Notice that what enters in the expression above is the part of $\M_L$ above the diagonal.

\subsection{Expressions for Energy}\la{energy}

We would like to compute the energy (\ref{EfromC}), repeated here for convenience
\beq
{\cal E}=-\phi^2 B(\lambda)\ ,\qquad B(g)=-\frac{1}{2}\sum_{a=1}^\infty{\mC_a\over\sqrt{1+{16g^2/ a^2}}}\,.
\eeq
For that, we need to find all $\mC_a$'s. That is done through the function $\rho(u)$.
Indeed, note that by rewriting \eq{re} using \eq{etafin} we have
\beq\la{rhofin}
{\rho(u)={\sum_a \pi\,{\mathbb C}_a\widehat K_a(u)\over\eta(u)} =\frac{\sum_a \pi\,{\mathbb C}_a\widehat K_a(u) }{\sum_a\frac{\pi\, {\mathbb C}_a}{c_a}K_a(u)-1}}\;.
\eeq
and therefore
\beq\la{rhotoe}
B(g)={i\over2}\lim_{u\to\infty}u\,\rho(u)\;.
\eeq
Moreover, from \eq{rhofin} we see that
\beq
\rho(ia/2)=c_a\;.
\eeq
Combining this with the definition of $c_a$ \eq{cdefinition} (and using that $\rho$ is an even function), we arrive at
\beq\la{eqa}
\frac{\rho(ia/2)}{2}=\frac{a}{2}+\int\limits_{-2g}^{2g}{du\over2\pi i}\frac{\rho(u)}{u-ia/2}\ ,\qquad a\in{\mathbb Z}^+\;.
\eeq
To better understand what this equation tells us we introduce the function
\beq\la{Hfunction}
H(u)=u+\int\limits_{-2g}^{2g}{dv\over2\pi }\frac{\rho(v)}{v-u}-\frac{i\rho(u)}{2}\;.
\eeq
This function has very familiar analytical properties. First, due to \eq{eqa}
it has zeroes at $v=ia/2$. Second, it also has poles at the zeroes of $\eta$, $\{u_k\}$, due to the $i\rho(u)/2$ piece. Finally, by its definition it seems like $H(u)$ has a
branch cut at the interval $[-2g,2g]$ due to the singularity of the integrand. However, the discontinuity of the integral, $i\rho$, precisely cancels with the discontinuity in $-i\rho/2$! Thus $H(v)$ is a meromorphic function with poles at zeros of $\eta$ and zeros at $ia/2$ (which are poles of $\eta$). As it behaves as $\sim u$ at infinity
we conclude that
\beq
H(u)=\frac{u}{\eta(u)}\;.
\eeq

We can now reconstruct $\rho(u)$ from $\eta(u)$. For $-2g<u<2g$ we can rewrite (\ref{Hfunction}) as
\beq\la{Hrho}
{u\over\eta(u)}-u=\dashint\limits_{-2g}^{2g}{dv\over2\pi }\frac{\rho(v)}{v-u}\;.
\eeq
Finding $\rho$ is now reduced to a standard Riemann-Hilbert problem. Using that $H(v)-v$ decays at infinity, we get for $u$ inside the cut
\beq\la{rhofinal}
\rho(u)=4\dashint\limits_{-2g}^{2g}\frac{dv}{2\pi}\frac{\sqrt{4g^2-u^2}}{\sqrt{4g^2-v^2}}\frac{1}{u-v}\[\frac{v}{\eta(v)}-v\]\ ,\qquad-2g<u<2g\;.
\eeq
Analytical continuation outside the cut is
\beq
\rho(u)=\int\limits_{-2g}^{2g}\frac{2dv}{\pi i}\frac{\sqrt{u^2-4g^2}}{\sqrt{4g^2-v^2}}\frac{v}{u-v}\frac{1-\eta(v)}{\eta(v)}+2iu\frac{1-\eta(u)}{\eta(u)}\;.
\eeq
Using (\ref{rhotoe}), we find for the energy
\beq\la{Bg}
\boxed{B_L(g)=\int\limits_{-2g}^{2g}{dv\over\pi}\frac{v^2}{\sqrt{4g^2-v^2}}\frac{1-\eta(v)}{\eta(v)}-\lim_{u\to\infty }u^2\frac{1-\eta(u)}{\eta(u)}}\;.
\eeq
At this point we can state the complete analytical solution of the initial problem of finding $B_L(g)$ for any given $L$.
In the next section we use this general solution to find a simple explicit expression for $B_L(g)$.

\subsubsection{Closed Expression for Energy}\la{Explicatesec}
Let us begin by evaluating the second term in \eq{Bg}. For that we have to expand $\eta(u)$ at large $u$, where $x_u$ is also large.
Using \eq{TtpP} and \eq{etaQQ2} we have
\beq
\nn\eta(u)
=\frac{u\[\sinh(2\pi u)P_L(x)-(-1)^L x^{-L-1}{\bf Q}(1/x)\]{\bf Q}(1/x)}{g\, C_1\,  x^{L+1} \sinh(2\pi u)}
\simeq \frac{u\, P_L(x){\bf Q}(1/x)}{g\, C_1\,  x^{L+1}}
\eeq
where the second step we drop the exponentially small terms.
Next using that at large $x$
\beq
\nn P_L(x)\simeq C_1 x^L\(1 + {C_2/C_1\over x^{2}}\)\ ,\qquad u= g\, x\(1+{1\over x^2}\)\ ,\qquad {\bf Q}(1/x)\simeq1+{f_{L+3}\over x^2}
\eeq
where
\beq
f_j=\sum_{i=1}^{L+1}C_{i}\,I_{j-L-2+2i}\;
\eeq
are the coefficients in the Laurent expansion of $P(x)\sinh(2\pi u)$. Putting the pieces together, we find
\beq
\eta(u)\simeq 1+\frac{1}{x^2}\(1+f_{L+3}+\frac{C^L_2}{C^L_1}\)
\quad\Rightarrow
\quad
-\lim_{u\to\infty }u^2\frac{1-\eta(u)}{\eta(u)}={g^2}\(1+f_{L+3}+\frac{C^L_2}{C^L_1}\)\;.
\eeq

We now turn to the integral in \eq{Bg}. We write it as a contour integral
\beq\la{integ}
\int\limits_{-2g}^{2g}{dv\over\pi}\frac{v^2}{\sqrt{4g^2-v^2}}\frac{1-\eta(v)}{\eta(v)}=
-g\oint\frac{dx}{2\pi i}\(1+\frac{1}{x^2}\)\frac{u}{\eta(u)}-2 g^2\;.
\eeq
In this form it is rather pointless to try to compute this integral by poles since there are infinitely many poles at $u=u_k$ situated roughly along the imaginary axis. Instead, the idea is to use
again \eq{TtpP} and \eq{etaQQ2} to factorize this function into two parts analytical inside/outside the unit circle
\beq\la{split}
\frac{u}{\eta(u)}=
\frac{g C_1}{x^{L+1}{\bf Q}(x)P_L(x)}
+
\frac{(-1)^Lg C_1}{1/x^{L+1}{\bf Q}(1/x)P_L(x)}\;.
\eeq
Note first that each of these terms separately has poles due to the zeros of $P_L(x)$ in the denominator. These poles, $\{y_i\}$, are localized on the unit circle (see appendix \ref{OrthogonalPolynomials} for a proof). They cancel between the two terms, so we can choose the integration contour in (\ref{integ}) to go slightly outside the unit circle. Second,
at least at small enough coupling, ${\bf Q}(x)$ does not have zeros inside the unit circle. That can be seen by noting that at weak coupling $\eta\to1$ and therefore, as follows from (\ref{etaQQ2}), the zeros of ${\bf Q}(x)$ can only be near $u=i a/2$.
Other values of the coupling are defined by analytic continuation. Now, the second term in (\ref{split}), analytical outside the unit circle contributes only by a residue at infinity. For the first term we contract the contour to the origin, picking the residue of the pole at the origin and the residues at the zeros of $P_L$. All in all, the integral in the r.h.s. of \eq{integ} gives\\
\beq
-g\oint\frac{dx}{2\pi i}\(1+\frac{1}{x^2}\)\frac{u}{\eta(u)}=2g^2\(1-f_{L+3}-\frac{C_2}{C_1}\)+g^2\sum_{i}\(1+\frac{1}{y_i^2}\)\frac{ C_1}{q(y_i)\d_y P_L(y_i)}\;.
\eeq
where we used that $q(y_i)=y_i^{L+1}{\bf Q}(y_i)$, see discussion above (\ref{qfunction}) and appendix \ref{qappendix}. Combining all terms together, we arrive at
\beq\la{Banalytic}
\boxed{
B=
g^2\(1-f_{L+3}-\frac{C_2}{C_1}\)+g^2\sum_{i}\(1+\frac{1}{y_i^2}\)\frac{ C_1}{q(y_i)\d_y P_L(y_i)}}\;.
\eeq

In the next section we will further simplify this expression. We will write it in a form which does not require finding the roots of $P_L$.
Before doing so, let us consider two simple cases where $L=0$ and $L=1$.
\begin{itemize}
\item For $L=0$ this equations simplifies considerably
since $P_0(x)=1/I_1$ is just a constant, so the sum over zeros vanishes. Moreover, $C_2=0$ and thus
\beq
B_{L=0}=g^2(1-f_3)=g^2(1-I_{3}C_{1})=g^2\(1-\frac{I_{3}(4\pi g)}{I_1(4\pi g)}\)
\eeq
in agreement with the localization prediction \eq{bplanar}!
\item For $L>0$ there is no result available from the side of the localization so it is interesting to see what our equation gives.
For $L=1$ things are still extremely simple. $P_1(x)=\pi  g\frac{x^2-1}{x I_2}$ furthermore using that
\beq
\frac{C_2}{C_1}=-1\ ,\qquad f_{L+3}=2\frac{I_4}{I_2}\ ,\qquad
\sum_i\frac{y_i^2+1}{y_i^2}\frac{C_1}{q(y_i)\d_y P_L(y_i)}=2\frac{I_3-I_1}{I_1}
\eeq
we find
\beq
B_{L=1}=g^2\frac{I_1 I_3+I_1 I_5-2 I_3^2}{I_1^2-I_1I_3}\;.
\eeq
\end{itemize}
We see that the result for $L=0$ and $L=1$ takes the form of a rational function of modified Bessel functions.
This holds true for any $L$, which is completely nontrivial from the expression \eq{Banalytic}.
This indicates that \eq{Banalytic} can be further simplified.

\subsubsection{Explicit Expression for Energy}

We will now simplify our result for the energy (\ref{Banalytic}) into an explicit combination of Bessel functions.

First, note that our general expression for $B_L$ has the form
\beq
B_L(g)=
g^2 A_L(g)+g^2 r_L(g)
\eeq
with
\beq
A_L\equiv 1-f_{L,L+3}-\frac{C_{L,2}}{C_{L,1}}\qquad\text{and}\qquad r_L\equiv \sum_{i}\(1+\frac{1}{y_i^2}\)\frac{ C_{L,1}}{q_L(y_i)\d_y P_L(y_i)}\;
\eeq
in this section we add an extra index $L$ to $C_i$ and $f_i$ to indicate that they are different for different $L$'s.

The key observation is that the complicated part $r_L$ can be expressed in terms of the simple part $A_i$
\beq\la{recursive}
r_L=2\sum_{i=0}^{L-1}(-1)^{L-i} A_i\qquad\Rightarrow\qquad B_L=
g^2 A_L+2g^2\sum_{i=0}^{L-1}(-1)^{L-i} A_i\;.
\eeq
We observed this relation from the explicit expressions for $A_L$ and $r_L$ for first several $L's$. We did not find any simple analytical
 proof of this relation for general $L$, however we verified the relation numerically with $200$ digits precision for $L=1,\dots,100$ for several values of $g$.
This relation allows us to overcome the step involving finding zeros of $P_L(y)$ which is rather hard to do analytically  for $L>3$.
Furthermore it allows to further simplify our result.
Next we use that
\beq
\frac{C_2}{C_1}=-\frac{\det\M^{(2,1)}_L}{\det\M^{(1,1)}_L}\ ,\qquad f_{L,L+3}=\frac{\det \M^{(1,2)}_{L+1}}{\det \M^{(1,1)}_{L+1}}=
\frac{\det \M^{(2,1)}_{L+2}}{\det \M^{(1,1)}_{L+2}}-1
\eeq
where $\M^{(a,b)}_L$ is the matrix obtained by deleting the $a^{\rm th}$ row and $b^{\rm th}$ column of $\M_L$.
The first equality is clear from \eq{PLex}. The second equality in the relation for $f_{L,L+3}$ holds true for any matrix whose entries
depend only on $|2i-2j+1|$. As a result we get
\beq
A_i=2-\frac{\det\M^{(2,1)}_{i+2}}{\det\M^{(1,1)}_{i+2}}+\frac{\det\M^{(2,1)}_{i}}{\det\M^{(1,1)}_{i}}\;.
\eeq
By plugging the $A_i$'s into (\ref{recursive}) we obtain our final result for the energy
\beq\la{BLsimplified}
\boxed{\boxed{
B_L(g)=g^2\(-\frac{\det\M^{(2,1)}_{L+2}}{\det\M^{(1,1)}_{L+2}}+2\frac{\det\M^{(2,1)}_{L+1}}{\det\M^{(1,1)}_{L+1}}-\frac{\det\M^{(2,1)}_{L}}{\det\M^{(1,1)}_{L}}\)
}}\;
\eeq
which can be also written in the form \eq{BLsimplified0}.

\section{Classical Limit}\la{sec:clas}
In this section we make an important test of our construction. We consider the classical limit $L\sim g\to\infty$
of our result \eq{BLsimplified} and compare it to the classical string prediction.
In the large $L$ limit our explicit result \eq{BLsimplified} becomes rather complicated since the size
of the matrix $M_L$ goes to infinity. We found it useful to rewrite \eq{BLsimplified} in terms of a
matrix model type of integral to perform this task. We also discuss the algebraic curve, arising both
in the classical open string and in the matrix model descriptions. We start from the string description and then move to the classical limit of our result.

\subsection{Classical Strings}
The classical string solution for general $L,\phi$ and $\theta$ is considered in details in appendix \ref{app:class}, where we derive its classical energy in a parametric form.
In particular for small angles it gives
\beqa\la{defOmega0}
L&=&4g\[{\mathbb K}\left(\omega^2\right)-{\mathbb E}\left(\omega^2\right)\]\;\\
E&=&L+g(\theta^2-\phi^2)\frac{1-\omega^2}{2{\mathbb E}(\omega^2)}
\eeqa
and therefore\footnote{In appendix \ref{strogapp} the small $L$ expansion of this result is compared with fixed $L$, large $g$ expansion of the TBA prediction. Despite the potential
 order of limits issues the results agree perfectly.}
\beq\la{BWS}
B^{\rm WS}=g\frac{1-\omega^2}{2{\mathbb E}(\omega^2)}\;.
\eeq

The energy (\ref{BWS}) is only one charge in an infinite set of conserved charges.
All these charges are nicely encoded in the so-called algebraic curve.
The algebraic curve is
 represented by the quasi-momentum $p(x)$ -- a function of the spectral parameter $x$ that encode all the conserve charges in its Taylor expansion \cite{Kazakov:2004qf}.
 In general, finding that function at strong coupling involves solving a linear system of differential equations.
Here we make a shortcut by relating our solution to ${\frak s \frak u}(2)$ Folded String, for which the algebraic curve is known \cite{Vicedo:2007rp,Beccaria:2011uz}.

\subsubsection{Relation to ${\frak s \frak u}(2)$ Folded String}
The $S^3$ part of the open string solution dual to the Wilson loop is analogous to the ${\frak s \frak u}(2)$ folded string.
To see the relation let us compare the global charges. For the folded string, these are the energy and
two angular momenta on $S^3$
\beqa
E^{\rm folded}&=&4 g  \sqrt{R^2+\frac{1}{R^2}-2 \cos (2 \phi )}\;\;{\mathbb K}\left(\sin ^2(\phi )\right)\\
J_1^{\rm folded}&=&{4 g \left(R+\frac{1}{R}\right) \[{\mathbb K}\left(\sin ^2(\phi )\right)-{\mathbb E}\left(\sin ^2(\phi )\right)\]}\la{J1folded}\\
J_2^{\rm folded}&=&{4 g \left(R-\frac{1}{R}\right) {\mathbb E}\left(\sin ^2(\phi )\right)}\;.
\eeqa
The result for the energy is given in a parametric form in terms of two parameters, $R$ and $\phi$ defined by the equations for $(J_1^{\rm folded},J_2^{\rm folded})$.
These parameters have a very clear meaning in terms of the algebraic curve.
The algebraic curve has two symmetric cuts with the branch points at $\pm R e^{\pm i\phi}$.
We immediately see that the expressions for $J_1^{\rm folded}$ (\ref{J1folded}) and $L$ (\ref{defOmega0}) are very similar under the identification
\beq
\omega=\sin\phi\;.
\eeq

In order to compare the folded string energy with the cusp solution we have to extract the $S^3$ contribution to the energy of the cusp solution in Appendix \ref{app:class}. This is given by\footnote{Which is $2g\, s$ in the notations of appendix \ref{app:class}.}
\beq
E_{S^3}=4g\,\omega\,{\mathbb K}({\omega^2})+{\cal O}(\theta^2)\;.
\eeq
We see that again the pattern of the elliptic functions is exactly the same and we have to relate the $\[4g\omega\]=\[4g\sin\phi\]$
pre-factor to $4g\sqrt{R^2+\frac{1}{R^2}-2 \cos (2 \phi )}$ that looks more complicated. However, if we set $R=1$
the square root simplifies to $\[8g\sin\phi\]$ and therefore, up to a factor of $2$, is exactly what we need.
Similar mismatch by $2$ appears in $J_1^{\rm folded}$ (and $J_2^{\rm folded}=0$). This factor of $2$ has a clear origin -- it is due to the fact that
the folded string is a closed string solution made of two folds, whereas the solution dual to the Wilson loop is an open string solution associated to a single fold. Having matched the parameters of the two solutions, we immediately read the $S^3$ part of the algebraic curve\footnote{The $AdS_3$ part should be identical to the $S^3$ part in order for their contributions to cancel in the BPS limit.}.

\subsubsection{Algebraic Curve}
\begin{figure}[ht]
\begin{center}
\includegraphics[scale=.35]{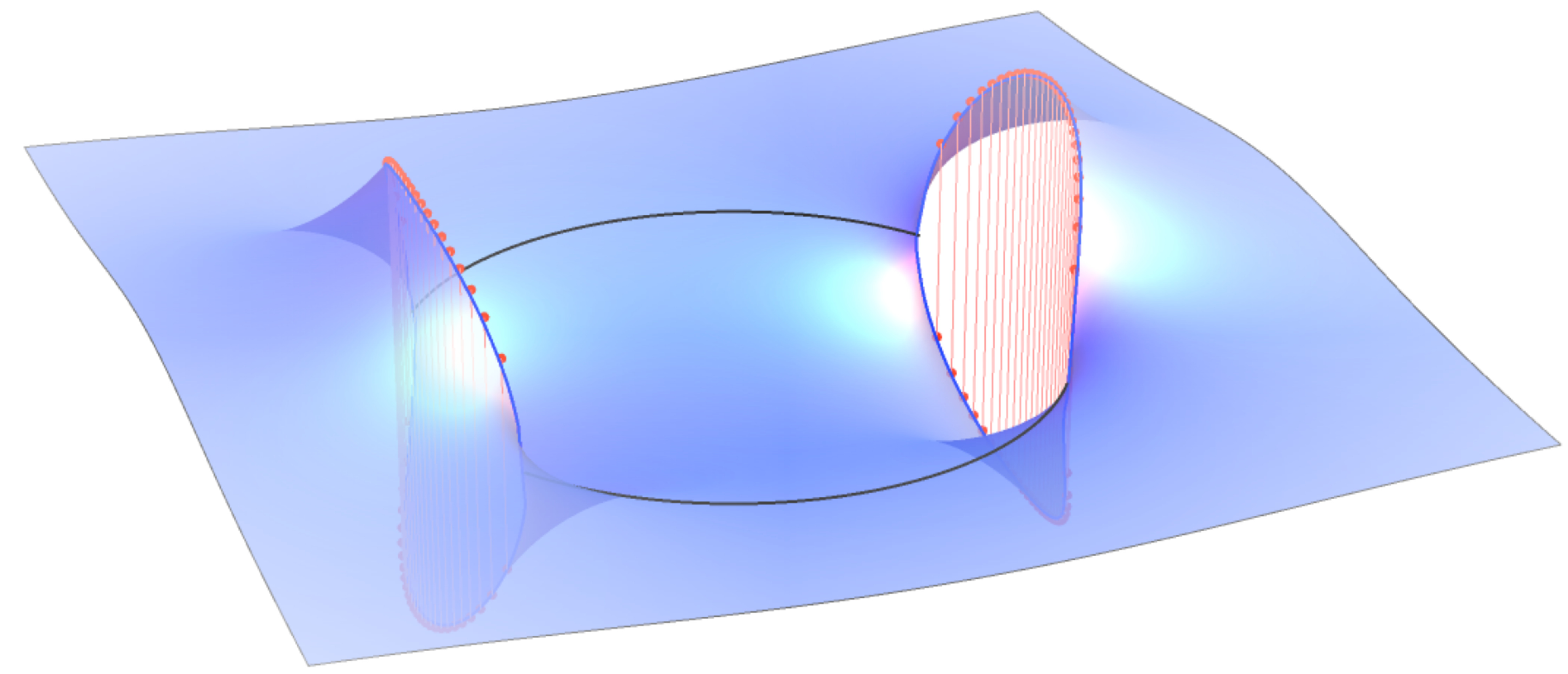}
\end{center}
\caption{\la{curve}\it Algebraic curve $\frac{x^2-1}{x}p(x)$ (real part is plotted). The cuts are discretized by zeros of $P_L(x)$ (red dots).}
\end{figure}

Above we identified uniquely the ($S^3$ part of) algebraic curve corresponding to our classical solution.
It should have two cuts on the unit circle with the branch points at $\pm e^{\pm i\phi}$ where $\phi=\arcsin\omega$\footnote{Curiously the $S^3$ part is precisely the Giant Magnon with momenta $\phi$.}.
This algebraic curve was studied in details in the relations to Giant Magnons, which are a particular limit of the
folded string solution. The quasimomentum is known explicitly
and for $|x|>1$ it reads
\beqa\la{pcl}
p(x)&=&\pi+2 {\mathbb K} \sqrt{\frac{1-x^2 e^{2 i \phi }}{-x^2+e^{2 i \phi }}} \left(\frac{2 i x \sin\phi}{x^2-1}-1\right)
+\frac{4  {\mathbb E} }{\cos\phi}F_1
-4\cos\phi\, {\mathbb K}\, F_2\\
\nn F_1&=&i{\mathbb F}\[
\sin ^{-1}\sqrt{\frac{\left(e^{2 i \phi }+1\right) \left(e^{i \phi
   }-x\right)}{\left(e^{2 i \phi }-1\right) \left(e^{i \phi }+x\right)}}
;
-\tan^2\phi
   \]\\
\nn F_2&=&i{\mathbb E}\[
\sin ^{-1}\sqrt{\frac{\left(e^{2 i \phi }+1\right) \left(e^{i \phi
   }-x\right)}{\left(e^{2 i \phi }-1\right) \left(e^{i \phi }+x\right)}}
;
-\tan^2\phi
   \]
\eeqa
where ${\mathbb K}\equiv {\mathbb K}(\sin^2\phi),\;
{\mathbb E}\equiv {\mathbb E}(\sin^2\phi)$.
For $|x|<1$ one should use the relation $p(x)=p(1/x)$.

One can see that the function $p(x)$ indeed has two cuts (see figure \ref{curve}).
Usually the quantization of the classical limit consists of a discretization of the cuts by poles.
For example, in the spectral problem the cuts are collections of the Bethe roots which
become dense in the classical limit. In our case of the cusp anomalous dimension, it seems naively that there are no Bethe roots at all since we are dealing with Vacuum -- the reference state on top of which the traditional Bethe roots correspond to excitations. From this point of view the situation is very different.
However, in the next section we argue that the cuts are formed by roots of the polynomials $P_L$. These are the holes $y_i$ in the effective Baxter equation (\ref{Tdef}).

Our interpretation of this is that the standard asymptotic Bethe ansatz at the cusp is in a sense an effective Bethe ansatz build over a nontrivial Vacuum. That vacuum, by itself, is quantized and is characterized by an additional set of roots $y_i$ (zeros of $P_L$). These ``hidden" variables
 $y_i$ are not visible in the standard ABA equations. Their existence is represented by the boundary scattering matrix and their contribution to the energy
 only comes about at finite size
 corrections. As a result, in the standard ABA equations that information is wiped out. In particular the ABA energy is exactly zero.

Another important difference between the spectral problem and the situation here is the following. Here, even in the classical scaling limit, the standard ABA equations do not describe the system exactly. This is contrary to the spectral problem where the Bethe equations describe accurately the classical limit including all finite size effects. That
difference should disappear when the extra level of $y_i$ is included into the Bethe equations.
In particular, as we will see, the ground state energy is described accurately in this case. It would be interesting to write the full system of Bethe equations which include the $y_i$'s. In other words, it would be interesting to include the excitations on top of the vacuum into the effective Baxter equation.

We will now turn to the classical limit of our result (\ref{BLsimplified}).

\subsection{Classical Limit of TBA}
In this section we give an interpretation of the algebraic curve, discussed in the previous section,
in terms of the quantities arising in the exact quantum TBA description.
As we will see, the key object in the analytical solution of the TBA equations is the polynomial $P_L(x)$. In the classical limit, the roots of $P_L$ will condense and form the ``vacuum" curve.

\subsubsection{Matrix Model Reformulation}
In order to compute the classical limit of our final result \eq{BLsimplified}
it is very convenient to rewrite the determinants in a form of multiple integrals.
For that we use the following integral representation of the modified Bessel function
\beq\la{Besseltoxn}
I_n={(-1)^{n+1}}\oint\frac{dx}{2\pi i} {e^{-2\pi g(x+1/x)}}{x^{n-1}}
\eeq
then, for example, $\det M_{L-1}$ becomes
\beq\la{MLm1}
\det \M_{L-1}=
\oint\prod_k^{L}\frac{dx_k}{2\pi i} e^{-2\pi g(x_k+1/x_k)}
\left|
\bea{lllll}
x_1^{-2}&x_1^0&\dots&x_1^{2L-6}&x_1^{2L-4}\\
x_2^{-4}&x_2^{-2}&\dots&x_2^{2L-8}&x_2^{2L-6}\\
\vdots&\vdots& \ddots & \vdots& \vdots\\
x_{L-1}^{2-2L}&x_{L-1}^{4-2L}&\dots&x_{L-1}^{-2}&x_{L-1}^{0}\\
x_{L}^{-2L}&x_{L}^{2-2L}&\dots&x_{L}^{-4}&x_{L}^{-2}
\eea
\right|
\eeq
the determinant under the integral in the r.h.s. is related to a Vandermonde determinant
in a simple way
\beq\la{ML1}
\det \M_{L-1}=\oint\prod_k^{L}\frac{dx_k}{2\pi i x^{2i-2}_k}\Delta(x_i^2) e^{-2\pi g \sum\limits_k^{L}(x_k+1/x_k)}
\eeq
where
\beq
\Delta(x_i^2)\equiv \prod_{i<j}(x_i^2- x^2_j)\;.
\eeq
To bring the expression to the form of the partition function of a matrix model we have to get
rid of the $x_k^{2i-2}$ in the denominator.
For that we use that $\Delta$ is antisymmetric, so we can (anti)-symmetrize this factor to bring it to a Vandermonde determinant
\beq
\sum_{\sigma}(-1)^{|\sigma|}\prod_k x_{\sigma_k}^{-2k+2}=\frac{\Delta(x_i^2)}{\prod_k x_k^{2L}}
\eeq
so that\footnote{This ensemble also describes the $O(-2)$ model in constant magnetic field near the critical point \cite{Bourgine:2011qp}.}
\beq
\det \M_{L-1}=\oint\prod_k^{L}\frac{dx_k}{2\pi i\, x_k^{2L}}\frac{\Delta^2(x_i^2)}{L!}e^{-2\pi g \sum\limits_k^{L}(x_k+1/x_k)}\;.
\eeq
As it is common in the matrix models let us introduce the resolvent
\beq\la{GL}
G_{L}(x)\equiv
\frac{1}{\det \M_{L-1}}
\oint\prod_k^{L}\frac{dx_k}{2\pi i\, x_k^{2L}}\frac{\Delta^2(x_i^2)}{L!}\frac{1}{L}\sum_{k}^{L}\frac{x}{x^2-x^2_k}e^{-2\pi g \sum\limits_k^{L}(x_k+1/x_k)}\;.
\eeq
Note that $\det\M_{L}^{(1,1)}=\det \M_{L-1}$, so in order to get the generalized Bremsstrahlung function
we need $\det\M_{L}^{(2,1)}=\det\M_{L}^{(L,1)}$. This can be also expressed in the form of the integral \eq{ML1} with an extra $\frac{1}{x_{L}^2}$ factor
which in terms of the resolvent becomes
\beq\la{MoM}
\frac{\det\M^{(2,1)}_{L}}{\det\M^{(1,1)}_{L}}=-{L G_{L}'(0)}\;.
\eeq
So far we used no approximation.
In the next section we evaluate this quantity in the classical limit $g\sim L\to\infty$.

\subsubsection{Saddle Point and Algebraic Curve}\la{Matrixsaddelpoint}
In the large $L$ limit we have to find the saddle point for the integral \eq{GL}. The saddle point equation is
\beq\la{saddle}
-\pi g \frac{x_k^2-1}{x_k^2}+\sum_{j\neq k}\(\frac{1}{x_k-x_j}+\frac{1}{x_k+x_j}\)-\frac{L}{x_k}=0\;.
\eeq
At the saddle point where (\ref{saddle}) is satisfied, the resolvent is simply given by
\beq
G_L(x)\simeq G^{\rm cl}_L(x)\equiv\frac{1}{2L}\sum_{k=1}^L \(\frac{1}{x-x_k}+\frac{1}{x+x_k}\)\;.
\eeq
Let us now show that in the classical $L\sim g\to\infty$ limit, the solution to the saddle point equation is given by an algebraic curve
with two cuts. In fact we will see shortly that this is precisely the classical algebraic curve arising in the
classical string description discussed in the previous section.
We introduce the function
\beq\la{pcl2}
p(x)\equiv\frac{L}{g}\frac{x}{x^2-1}-\frac{2L}{g}\frac{x^2}{x^2-1}G_L^{\rm cl}(x)
\eeq
in terms of this function the equation \eq{saddle} simply becomes
\beq\la{Bethecon}
p(x_k+i0)+p(x_k-i0)=-2\pi\;.
\eeq
This is an equation familiar from the general classical algebraic curve approach \cite{Kazakov:2004qf}.
Furthermore one can see from \eq{saddle} that the set of roots $\{x_i\}$ is invariant under inversion. This implies that
\beq
p(x)=p(1/x)\;.
\eeq
Together with the other asymptotic properties
\beq
p(-x)=-p(x)\ ,\qquad p(x)\simeq -\frac{L}{gx}+{\cal O}(1/x^3)
\eeq
the function $p(x)$ is uniquely determined.
We can reconstruct the functions solely from these properties along the lines of \cite{Kazakov:2004qf},
however all these properties are exactly the same as the properties of $p(x)$ in \eq{pcl}!
So we conclude that $p(x)$ we defined in terms of the roots in \eq{pcl2} is the one written in \eq{pcl}.
This implies that we can immediately extract the resolvent and thus the ratio $\det\M_L^{(2,1)}/\det\M_L^{(1,1)}$
in \eq{MoM}
\beq
\!\frac{\det\M_L^{(2,1)}}{\det\M_L^{(1,1)}}\simeq
-{L G_{L}'(0)}{}=\lim\limits_{x\to 0}\d_x\[\frac{g \left(x^2-1\right) p(x)}{2
   x^2}-\frac{L}{2 x}\]=-\frac{L}{3}\[1+\omega^2\frac{{\mathbb E}(\omega^2)+{\mathbb K}(\omega^2)}{{\mathbb E}(\omega^2)-{\mathbb K}(\omega^2)}\]
\eeq
where $\omega$ is an implicit function of $\frac{L}{g}$ via ${\mathbb K}(\omega^2)-{\mathbb E}(\omega^2)=\frac{L}{4g}$.

Now having this expression computed we obtain the generalized Bremsstrahlung function from \eq{BLsimplified}
\beq
B^{\rm cl}=
-g^2 \d_L^2\frac{\det\M_L^{(1,2)}}{\det\M_L^{(1,1)}}
\eeq
using that $\d_L=\frac{1-\omega^2}{4g\omega {\mathbb E}(\omega^2)}\d_\omega$ we get
\beq
B^{\rm cl}=
g\frac{1-\omega^2}{2{\mathbb E}(\omega^2)}
\eeq
in the perfect agreement with \eq{BWS}!

\subsubsection{Zeros of $P_L$ as Quantization of Algebraic Curve}

In this section we will show that in the classical limit zeros of $P_L(x)$ coincide with the solution to the saddle point equation (\ref{saddle}). Therefore, they can be thought of as the quantization of the algebraic curve.

Note first that up to a constant factor, $P_{L-1}(x)$ is given by a determinant of Bessel functions and the monomials of $x$ (\ref{PLex}). Therefore, using (\ref{Besseltoxn}) it takes the same form as the matrix-model type of the integral (\ref{MLm1}). The only difference is that in the last line in \eq{MLm1} we write $x$ instead of $x_L$ and remove the integration over $x_L$. In the classical limit the integral is saturated by the saddle point
values $x_i\to x_i^{\rm cl}$ (\ref{saddle}), so that $P_{L-1}(x)$ becomes proportional to the determinant with fixed $x_i=x_i^{\rm cl}$. Now it is obvious to see that
the determinant is zero whenever the argument $x$ is equal to $\pm x_i^{\rm cl}$. We conclude that
in the roots of $P_L(x)$, denoted by $y_i$, satisfy the saddle point equation
\beq\la{saddle}
g\sim L\to\infty\;\;:\;\;\frac{{L}\,y_k }{y_k^2-1}-\frac{y_k^2}{y_k^2-1}\sum_{j\neq k}^{2L}\frac{1}{y_k-y_j}=n_k\pi \, g\ ,\qquad k=1,\dots,2L
\eeq
where $n_k={\rm sign}\;\RE y_k^\text{}$. It means that the zeros of $P_L(x)$ discretize the curve and allows us to interpret them as the quantization of the classical curve. At the same time, these zeros are the ``holes" in the effective Baxter equation \eq{TtpP}.

Note that \eq{saddle} formally coincides with the strong coupling limit of ${\frak s\frak u(2)}$ spectral Bethe ansatz in the case where the number of roots is equal to the length (which is equal to $2L$ due to the open/closed double folding)! This seems to suggest that the initial vacuum with $Z^L$ at the cusp as being a maximally excited state of some other reference state. It would be interesting to understand this point in more details.\footnote{Note that the choice of treating $Z^L$ as the cusp vacuum was very convenient as it preserve supersymmetry in the BPS configuration. Other possible choice is to use a complex scalar made of the same scalars that the Wilson lines couple to. Away from the BPS configuration, supersymmetry is anyhow broken and that other choice preserve more (bosonic) symmetry.}

Having established the classical limit of the zeros of $P_L$ (\ref{saddle}), we can now write an explicit expression for the classical limit of $P_L$ itself. That is done in appendix \ref{app:qcPL}.

An interesting question is whether there is a simple generalization of \eq{saddle} for any $L$ and $g$. For example, for $g=0$ and any $L$ we find that $P_L(x)$ is related to the Chebyshev polynomials (see appendix \ref{OrthogonalPolynomials}).
Their zeros are known explicitly $y_k=e^{i\pi k/L}$ and they satisfy an equation similar to \eq{saddle}
\beq\la{saddle2}
g=0,\;{\rm any}\;L\;\;:\;\;\frac{L+1/2}{y_k}-\sum_{j\neq k}^{2L}\frac{1}{y_k-y_j}-\frac{2y_k}{y_k^2+1}=0\ ,\qquad k=1,\dots,2L\;.
\eeq
Note that (\ref{saddle2}) is similar to the $SU(2)$ case \cite{Gromov:2007ky}.

\section{Summary and Discussion}
In this paper, we have computed analytically the quark--anti-quark potential on $S^3$ in the near BPS configuration and with $L$ units of R-charge. Let us summarize the main steps in the derivation and identify potentially important future directions.

We started from the vacuum Thermodynamic Bethe Ansatz (TBA) equations in the so-called ``simplified" form listed in section~\ref{sec:BremsstrahlungTBA}.
By using the methods developed for the spectral problem (known as FiNLIE) \cite{talk,Gromov:2011cx} we reduced this infinite set of integral equations to just three equations for three functions $\rho(u),\eta(u)$ and $\varrho(u)$. Quite remarkably,
only their values on the cut $u\in [-2g,2g]$ enter the equations, making them very convenient
for numerics. By solving them numerically for various values of the coupling, we confirmed with high accuracy
the localization result for $L=0$ and made numerical predictions for other values of $L$. Next, by using assumption about the simplicity of the analytical properties of $\eta$, $\rho$ and by analyzing the structure of their poles and zeroes, we derived an explicit equations for them (\ref{etaQQ2},\ref{rhofinal}). These equations agree nicely with the results of our numerics, verifying the assumption!

Perhaps, even more excitingly we found that the zeros of $\eta$ (or equivalently the poles of $\rho$) satisfy some kind of Baxter equation \eq{Tdef}.
This Baxter equation has a rather curious form -- instead of the familiar quantum shifts in the spectral parameter $u$ by $\pm  i/2$, we observe a ``crossing" type of shifts $x\to 1/x$. Curiously, the same shift emerged before in the strong coupling Thermodynamic Bubble Ansatz (TBuA) equations for {\it classical} minimal surfaces in AdS
\cite{Alday:2009dv,Alday:2010vh, students}.
Our construction  on the other hand is completely {\it quantum}. That is to say, it is valid at any value of the coupling. It would be very interesting to see if our approach provides a natural quantization of the minimal surfaces linking these two approaches.

Another important object, emerging in our consideration, is the polynomial $P_L(u)$. Zeroes of $P_L(u)$ are the ``holes", i.e. nontrivial zeroes of the Baxter function ${\bf T}(u)$.
We have found that the polynomial $P_L(u)$ is related to a matrix model resolvent.
This interpretation was especially useful in the classical limit $L\sim g\to \infty$ where the matrix model integral is saturated by a saddle point.
Very naturally, the algebraic curve describing the saddle point of the matrix model coincides precisely with the classical algebraic curve of the integrable worldsheet theory.
Moreover, we have shown that the roots of $P_L(u)$ coincide with the matrix model eigenvalues in the classical limit. This implies that these roots provide a quantization of the classical algebraic curve.

There are several interesting future directions of study:
\begin{itemize}
\item It would be interesting to understand the physical interpretation of the matrix model we obtained. Note that this matrix model is different from the ones usually arising from localization techniques where one is dealing with $N_c\times N_c$ matrices. Here instead, we are working in the planar limit from the very beginning and the matrix model we obtain is for $L\times L$ matrices, where $L$ is the number of scalar $Z$ fields inserted at the cusp.
One possible interpretation is that the matrix model is related to a quantization of the transfer matrix in the physical channel. Such picture connects neatly with the TBuA for minimal surfaces discussed above
where the elements entering into the equations are the transfer matrices between edges of a polygon Wilson loop.

\item Whereas for $L=0$ there is already known result from localization (which we successfully reproduced),
for $L>0$ our results are new. The localization techniques of \cite{Correa:2012at,Fiol:2012sg,localizations,Giombi:2012ep} can be generalized to the case considered here.\footnote{We thank S. Giombi for discussion on this point.} Our results therefore provide for the first time a prediction from integrability for the localization techniques. Having an alternative derivation of the results may shed some light on the relations between the two very different methods.

\item It may be interesting to find a similar regime of the spectral TBA to the one considered here. For example,
the slope function \cite{Basso:2011rs} in ABJM theory \cite{Bombardelli:2009xz,Gromov:2009at} is one of possible candidates. The ``exceptional operators" recently discussed in \cite{Arutyunov:2012tx} is another.

\item In this paper we considered the near BPS configuration in the case where the angles, $\phi$ and $\theta$ are small. It should not be too hard to generalize our result to the case where $(\phi^2-\theta^2)$ is small but $\phi$ is finite. In particular, it would be nice to understand the generalization of our classical curve.\footnote{We thank N.~Drukker for discussion on this point.}

\item It would be interesting to compare our results at strong coupling with the string computation at one loop. From the TBA point of view this would require solving the matrix model to the next order (a part of the result is in appendix \ref{strogapp}). On the classical string side of the calculation, it should be possible to obtain the one loop result from the algebraic curve \cite{Gromov:2007aq}.
As we already know the algebraic curve, the result to a large extent can be read from \cite{Gromov:2008ec,Gromov:2011de,Beccaria:2011uz} (up to a subtlety of open vs. closed strings).

\item We hope that the methods developed here could be extended to any value of the angles $\phi$, $\theta$ and in particular to the general spectral TBA. The analog of the function $\eta$ with simple analytic properties should also exist in the general spectral TBA. The properties we observed indicate that this function is an important object in general.

\item In this paper we concentrated on the vacuum energy of the flux between the quarks. Introducing excitations may lead to an interesting interaction between the conventional Bethe Ansatz and the matrix model.

\item Finally, it would be interesting to solve the cusp TBA in other special limits of the parameters. For example, in the ladders limit where $i\theta\to\infty$ with $g\,e^{i\theta/2}$ fixed, the TBA should reduce to a Schrodinger problem \cite{Correa:2012nk}.
\end{itemize}

We hope to come back to these points in the future.

\subsection*{Acknowledgments}
We would like to thank D.~Correa, N.~Drukker, S.~Frolov, S.~Giombi, I.~Kostov, J.~Maldacena, J.~Penedones, P.~Vieira and D.~Volin for discussions.
We are especially grateful to F.~Levkovich-Maslyuk and P.~Vieira for careful reading of the manuscript before publication.
N.G. is especially indebted to V.~Kazakov for his lectures on matrix models.
N.G. (A.S.) would like to thank Perimeter Institute (King's College London, LPT ENS and CEA Saclay) for warm hospitality.
Research of N.G. is partly supported by STFC grant ST/J002798/1. Research at the Perimeter Institute is supported in part by the Government of Canada through NSERC and by the Province of Ontario through MRI. The research of A.S. has been supported in part by the Province of Ontario through ERA grant ER 06-02-293 and by the U.S. Department of Energy grant \#DE-FG02-90ER4054.

\appendix

\section{Kernels conventions}\la{kernelsconventions}

For the TBA in section \ref{sec:BremsstrahlungTBA} we use the standard definitions
\beq
f^{[a]}(u)=f(u+ia/2)\ ,\qquad{\frak s}(u) = \frac{1}{2\cosh(\pi u)}\ ,\qquad I_{m,n}=\delta_{m+1,n}+\delta_{m-1,n}\;.
\eeq
Our conventions for the kernels are
\beqa
 \widetilde K_{ab}&=&{\cal R}^{(10)}_{ab}+{\cal B}^{(10)}_{ab-2}= \frac{1}{2}\(\widetilde K_a^{[+b-1]}-\widetilde K_a^{[-b+1]}+
K_a^{[+b-1]}+K_a^{[-b+1]}\)+\sum_{r=1}^{a}K_{b-a-3+2r}\nn\\
&&{\cal R}_{2n}^{(01)}=\frac{1}{2}\(\widehat K_n^+-\widehat K_n^-+K_n^++K_n^-\)\\
\widehat K_{ba}&=&{\cal R}_{ba}^{(01)}+{\cal B}_{b-2,a}^{(01)}= \frac{1}{2}\(
\widehat K_a^{[+b-1]}-\widehat K_a^{[-b+1]}+K_a^{[+b-1]}+K_a^{[-b+1]}\)+\sum_{r=1}^{a}K_{b-a-3+2r}\nn
\eeqa
where $K_a$, $\widehat K_a$ and $\widetilde K_a$ are given in (\ref{thekernels}). The functions ${\cal R}$ and ${\cal B}$ are the standard ones defined in \cite{Gromov:2009bc} and are not used
in the main text.

For convenient, we summarize below the notations of reference \cite{Correa:2012hh} used in section \ref{sec:BremsstrahlungTBA}
\beq
\begin{array}{lcl}
\Psi&=& -(1+Y_{1,1})/\phi^2\\
\Phi&=&-(1+1/Y_{2,2})/\phi^2\\
\cY_m&=&Y_{m,1}=1/Y_{1,m}\\
{\cal X}_m&=&-(1-1/(Y_{1,m}Y_{m,1}))/\phi^2\\
{\mathbb C}_a^2&=&\lim_{\substack{u\to0}}\(4u^2 Y_{a,0}\)/\phi^4\\
{\cal E}&=&-\phi^2 B(\lambda)
\end{array}
\qquad\text{at}\quad\phi\to0
\eeq
and the new notations introduced throughout this paper
\beq\la{summarynotations}
\begin{array}{lcl}
{\cal Y}_m&=&\frac{{\cal T}_{m}^+{\cal T}_{m}^-}{{\cal T}_{m+1}{\cal T}_{m-1}}-1\\
{\cal T}_m&=&m+K_m\hat *\rho\\
{\mathbb Y_m}&=&\frac{{\mathbb T}_{m}^+{\mathbb T}_{m}^-}{{\mathbb T}_{m+1}{\mathbb T}_{m-1}}-1= {{\cal Y}_m}(1+\phi^2{{\cal X}_m})\\
{\mathbb T}_m&=&{\cal T}_m+\phi^2 \tau_m\\
\tau_m&=&-\frac{m^3}{12}+mu^2+K_m\hat*\varrho+\sum\limits_{n=-\infty}^\infty\!\! b_{n} K_{m-n}\\
\eta&=&\frac{\Psi\,{\cal T}_2}{{\cal T}^{-_+}_1\,{\cal T}_1^{+_-}}=\frac{\Phi\,{\cal T}_2}{{\cal T}^{-_-}_1\,{\cal T}_1^{+_+}}
\end{array}
\qquad\quad
\begin{array}{lcl}
c_0&=&\rho(0)\\
c_{m>0}&=& \cT_m(0)\\
y_i&=&\text{zeros of }P_L\\
x_k&=&\text{zeros of }\eta\\
y_a&=&x(ia/2)
\end{array}
\eeq

\section{Derivation of $q$}\la{qappendix}
In this appendix we give some more details about the function $q$ defined in section \ref{Qfunctionsec} and used in section \ref{Explicatesec}.

The function $y^{L+1}\bQ(y)$ is defined as the sum of positive powers of $y$ in the Laurent expansion of $P_L(y)\sinh(2\pi u)$, see (\ref{Tdef}), (\ref{eqTlarge}). It is therefore almost proportional to $P_L(y)$. That is, we have
\beq
y^{L+1}\bQ(y)=q(y)+P(y)f(y)
\eeq
where $q(y)$ has a finite Laurent expansion and $f(y)$ is some function with a regular expansion. One can always shift $q$ by another function proportional to $P_L$. The minimal choice of $q$ is such that it has Laurent degree $L+1$. It is given by
\beq\la{defq}
q(y)\equiv \[P(y)\sinh(2\pi u)\]_+-P(y)\[\sinh(2\pi u)\]_+=\!\!\oint\limits_{|x|=|y|}\!\!{dx\over 2\pi i}{x \sinh(2\pi u_x)\over x^2-y^2}\[P(x)-P(y)\]\;.
\eeq
It follows from its definition that at the zeros of $P_L$ we have $q(y_i)=y_i^{L+1}\bQ(y_i)$. Moreover, for any $y$ it satisfies
\beqa
q(1/y)\!\!&=&\!\!(-1)^L \[P(y)\sinh(2\pi u)\]_--(-1)^LP(y)\[\sinh(2\pi u)\]_-\\
\!\!&=&\!\! (-1)^L\(P(y)\sinh(2\pi u)-\[P(y)\sinh(2\pi u)\]_+\)-(-1)^LP(y)\(\sinh(2\pi u)-\[\sinh(2\pi u)\]_+\)\nn\\
\!\!&=&\!\!(-1)^{L+1}q(y)\nn\;.
\eeqa
By this we prove that on the roots of $P_L$ the function $y^{L+1}{\bf Q}$ can be replaced by a simpler function $q(u)$.

\section{Orthogonal Polynomials}\la{OrthogonalPolynomials}

In these appendix we relate the functions $P_L$ to orthogonal polynomials with a simple measure. We use this fact to prove that the zeros of $P_L(x)$ are located on the unit circle. This
was also used in the derivation of a closed expression for the energy in section \ref{Explicatesec}.

We defined the coefficients in the Laurent expansion of $\bT(y)$ as
\beq\la{fss}
f_s\equiv-\oint{dy\over2\pi i\, y}\frac{\sinh(2\pi u)P_L(y)}{y^{s}}=-\frac{1}{2}\oint{dy\over2\pi i\, y} {\sinh(2\pi u)P_L(y)}
\(\frac{1}{y^{s}}+(-1)^Ly^{s}\)\;.
\eeq
Using the explicit form for $P_L(y)$ \eq{PLex} we can evaluate this integral in terms of the coefficients $C_i$
\beq
f_j=\sum_{i=1}^{L+1}C_{i}\,I_{j-L-2+2i}\;.
\eeq
It follows from \eq{gap} $f_s=0$ for $s={0,\dots,L}$ and $f_{L+1}=1$.
Let us change the variable of integration to
 $z\equiv \frac{1}{2}\(y+1/y\)=\frac{u}{2g}$. The integration measure in (\ref{fss}) transforms as
\beqa
\(\frac{1}{y^{s}}+(-1)^Ly^{s}\){dy\over y}=
\left\{
\bea{ll}
+\frac{2^s z^s+\dots}{\sqrt{z^2-1}} dz\;\;&,\;\;L\text{ is even}\\
-(2^s z^{s-1}+\dots) dz\;\;&,\;\;L\text{ is odd}
\eea
\right.
\eeqa
Let us consider separately the case where $L$ is even and odd.

\paragraph{Even $L$.}
We define $T^g_{L+1}(z)\equiv 2\pi g z P_L(x(z))$. It is a polynomial of odd degree $L+1$ in $z$ variable which obeys
\beq
\int\limits_{-1}^1 dz\,\mu(z)\,
T_{n}^g(z)\,
z^{s} =
\left\{
\bea{cl}
\frac{\pi}{2^n}&,\;\;s=n\\
0&,\;\;s<n
\eea
\right.\qquad\text{where}\qquad\mu(z)=\frac{\sinh(4\pi g z)}{4\pi g z\sqrt{1-z^2}}\;.
\eeq
At weak coupling $g\to 0$ the measure reduces to $\frac{1}{\sqrt{1-z^2}}$ and $T_n^g$ becomes a
 Chebyshev polynomial of the first kind $T_n$.

\paragraph{Odd $L$.}
Note that for odd $L$ we have $P_L(\pm 1)=0$ and therefore we can always extract the factor $y-1/y=2\sqrt{z^2-1}$.
We define $U^g_{L}(z)\equiv 2\pi g z\frac{P_L(x)}{\sqrt{z^2-1}}$. It is a polynomial of odd degree $L$ in $z$ variable which obeys
\beq
\int\limits_{-1}^1dz\,\mu(z)\,
U_{n}^g(z)\,
z^{s} =
\left\{
\bea{cl}
\frac{\pi}{2^{n+1}}&,\;\;s=n\\
0&,\;\;s<n
\eea
\right.\qquad\text{where}\qquad\mu(z)=\frac{\sinh(4\pi g z)}{4\pi g z}\sqrt{1-z^2}\;.
\eeq
At weak coupling $g\to 0$ the measure reduces to ${\sqrt{1-z^2}}$ and $U_n^g$ become a
 Chebyshev polynomial of the second kind $U_n$.

 The relation to the orthogonal polynomial is quite remarkable because this automatically implies that
 all zeros of $P_L(x)$ can only appear on the unit circle. This follows from the fact that all zeros of orthogonal polynomials on the interval $[-1,1]$
must be inside the interval for a positive defined measure. Indeed, let's assume that only some of zeroes of
an orthogonal polynomial $p_n(z)$ are real and denote them $z_n$, then the function $p_n(z)\prod_i(z-z_i)$
is either always positive or negative on the interval. This is in contradiction with the orthogonality which
tells that the integral over the interval should vanish. The image of the interval $[-1,1]$ under the map $\frac{1}{2}(x+1/x)=z$ is the unit circle which doubles the number of roots.

\section{Quasi-classical $P_L$}\la{app:qcPL}
Knowledge of the quasimomemtum $p(x)$ and as a result of the resolvent $G(x)$
allows us to find the polynomial $P_L(x)$ in the classical limit.
In view of importance of this quantity we make this calculation here.
This calculation mainly consists in the computing of an integral of quasimomentum $p(x)$
\beqa
&&g\int\limits_{x}^\infty d\tilde x\(1-\frac{1}{\tilde x^2}\)p(\tilde x)  =
-g\int\limits_{x}^\infty d\tilde x\(\tilde x+\frac{1}{\tilde x}\)p'(\tilde x)  -{L}
-g\frac{x^2+1}{x}p(x)
\eeqa
where we integrated by part.
 The reason is that $p'(x)$ is much simpler then $p(x)$ itself
and does not contain any elliptic functions.
As a result the integral can be computed explicitly
\beqa
&&g\int\limits_{x}^\infty d\tilde x\(\tilde x+\frac{1}{\tilde x}\)p' (\tilde x)=
\frac{4  g \mathbb{K} \sqrt{y^2-x^2} \sqrt{1-x^2 y^2}}{\left(x^2-1\right) y}-4 g
   \mathbb{K}+L \log (\Lambda )\\&&-\frac{L}{2} \log \left(\frac{2 x^4 y^2-x^2
   \left(y^2+1\right)^2+2 y^2}{4 x^2 y^2}+\frac{\left(x^2-1\right) \sqrt{y^2-x^2}
   \sqrt{1-x^2 y^2}}{2 x^2 y}\right)\nn
\eeqa
the integral in fact is log-divergent and we introduce a cut-off $\Lambda$.
We also denote $y=e^{i\phi}$ is the branch point.
At the same time from the relation to the resolvent the integral gives
\beq
g\int\limits_{x}^\infty  d\tilde x\(1-\frac{1}{\tilde x^2}\)p(\tilde x) =\log\[\prod_i^{2L} \frac{x-x_i}{\sqrt x}\]-L\log\Lambda
\eeq
and we see that the divergent part cancels. Since $\log P^{\rm cl}=\log\[C_1\prod_i^{2L} \frac{x-x_i}{\sqrt x}\]$
and $C_1$ can be found in the classical limit to be $C_1=\exp\[-L\log\frac{\(y-1/y\)^2}{4}-8g\mathbb E\;\]$ we finally get
\beqa
\log P^{\rm cl}(x)&=&-4g{\mathbb E}
-g\frac{x^2+1}{x}p(x)+L\log \frac{2y}{y^2-1}
-\frac{4  g \mathbb{K} \sqrt{y^2-x^2} \sqrt{1-x^2 y^2}}{\left(x^2-1\right) y}\\
&+&\frac{L}{2} \log \left(\frac{2 x^4 y^2-x^2
   \left(y^2+1\right)^2+2 y^2}{ x^2 (y^2-1)^2}+2\frac{\left(x^2-1\right)y \sqrt{y^2-x^2}
   \sqrt{1-x^2 y^2}}{ x^2 (y^2-1)^2}\right)\nn\;
\eeqa
where again $y=e^{i\phi}$.

\section{Classical String}
In this appendix we solve the classical string problem at any value of $L$ and the angles $\phi$, $\theta$. The solution is given in parametric form in terms of elliptic functions. We then expand the general result in various limits.

\la{app:class}
\subsection{Solution for Arbitrary Angles and $L$}
We are following \cite{Correa:2012hh} extending these results to the most general case with arbitrary $L$, $\phi$ and $\theta$.
We start from the following ansatz for the embedding coordinates in $AdS^3$:
\beq
y_1+iy_2=e^{i\kappa\tau}\sqrt{1+r^2(\sigma)}\ ,\qquad
y_3+iy_4=r(\sigma) e^{i\varphi(\sigma)}
\eeq
and $S^3$:
\beq
x_1+i x_2=e^{i\tau\gamma}\sqrt{1-\rho^2(\sigma)}\ ,\qquad
x_3+i x_4=\rho(\sigma)e^{i f(\sigma)}\;.
\eeq
We use the conformal gauge with energy momentum tensor normalized to $1$. In this gauge the
range of the worldsheet coordinate $\sigma\in[-s,s]$ is nontrivial and has to be found dynamically.
The ends of the strings should go to the boundary in $AdS_3$ which implies $r(\pm s/2)=\infty$.
Furthermore the angles are defined as $\varphi(\pm s/2)=\pm(\pi-\phi)/2$
and $f(\pm s/2)=\pm\theta/2$. In order for the ansatz to satisfy the equations of motion and the Virasoro constraint, the functions $\phi,\varphi$ and $\rho,r$ should satisfy the following equations
\beqa
\ell_\theta&=&-\rho^2 f'\\
\frac{\rho^2(\rho')^2}{1-\rho^2}&=&-\ell_\theta^2-(\gamma^2-1)\rho^2+\gamma^2\rho^4\equiv D_\theta\\
\ell_\phi&=&r^2\varphi'\\
\frac{r^2(r')^2}{1+r^2}&=&-\ell_\phi^2+(\kappa^2-1)r^2+\kappa^2r^4\equiv D_\phi\;.
\eeqa
From these equations we can write the angles and charges directly, avoiding solving the equations themselves. We have
\beqa
s&=&\int_{-s/2}^{s/2} d\sigma = 2\int_{r_0}^\infty dr\frac{r}{\sqrt{1+r^2}\sqrt {D_\phi}}=\int_{\rho_0}^1d\rho\frac{2\rho}{\sqrt{1-\rho^2}\sqrt D_\theta}\\
\phi-\pi&=&\int_{-s/2}^{s/2}\varphi' d\sigma=-2\int_{r_0}^\infty dr \frac{\ell_\phi}{r\sqrt{1+r^2}\sqrt{D_\phi}}\\
\theta&=&\int_{-s/2}^{s/2}f' d\sigma=2\int_{\rho_0}^1d\rho\frac{\ell_\phi}{\rho\sqrt{1-\rho^2}\sqrt D_\phi}\\
E&=&\[4g\kappa\int_{r_0}^\infty d r\frac{r\sqrt{1+r^2}}{\sqrt D_\phi}\]_{\rm finite\;part}
\\
L&=&4g\gamma\int_{\rho_0}^1d\rho\frac{\rho\sqrt{1-\rho^2}}{\sqrt D_\theta}
\eeqa
where $\rho_0\;$ ($r_0$) is a solution to the equation $D_\theta=0\;$  ($D_\phi=0$).
The ``finite part" means that the integral is power-like divergent and one should keep only the finite part.
These integrals are standard elliptic integrals and give
\beqa
s&=&\frac{2 \sqrt{2} }{\sqrt{\gamma ^2+k_\theta^2+1}}{\mathbb K}\left(\frac{-k_\theta^2+\gamma ^2+1}{k_\theta^2+\gamma ^2+1}\right)
=\frac{2 \sqrt{2} }{\sqrt{\kappa^2+k_\phi^2+1}}{\mathbb K}\left(\frac{-k_\phi^2+\kappa^2+1}{k_\phi^2+\kappa^2+1}\right)
\\
\nn\theta&=&\frac{2\ell_\theta}{k_\theta
   \left(1+k_\theta^2-\gamma ^2\right)}\[{\left(1+\gamma ^2+k_\theta^2\right) \Pi \left(\frac{k_\theta^2-2 \ell_\theta^2-\gamma ^2+1}{2 k_\theta^2} \Big|\frac{k_\theta^2-\gamma ^2-1}{2
   k_\theta^2}\right)}\right.\\\nn&-&\left.{2 \gamma ^2 {\mathbb K}\left(\frac{k_\theta^2-\gamma ^2-1}{2 k_\theta^2}\right)}\]
\\
\nn\phi&=&\frac{2\ell_\phi}{k_\phi
   \left(1+k_\phi^2-\kappa ^2\right)}\[{\left(1+\kappa ^2+k_\phi^2\right) \Pi \left(\frac{k_\phi^2-2 \ell_\phi^2-\kappa ^2+1}{2 k_\phi^2} \Big|\frac{k_\phi^2-\kappa ^2-1}{2
   k_\phi^2}\right)}\right.\\\nn&-&\left.{2 \kappa ^2 {\mathbb K}\left(\frac{k_\phi^2-\kappa ^2-1}{2 k_\phi^2}\right)}\]\\
\nn L&=&-2 \sqrt{2} g \frac{\sqrt{\gamma ^2+k_\theta^2+1}}{\gamma }\left[{\mathbb E}\left(\frac{-k_\theta^2+\gamma ^2+1}{k_\theta^2+\gamma
   ^2+1}\right)-{\mathbb K}\left(\frac{-k_\theta^2+\gamma ^2+1}{k_\theta^2+\gamma ^2+1}\right)\right]\\
\nn E&=&-2 \sqrt{2} g \frac{\sqrt{\kappa ^2+k_\phi^2+1}}{\kappa }\left[{\mathbb E}\left(\frac{-k_\phi^2+\kappa ^2+1}{k_\phi^2+\kappa
   ^2+1}\right)-{\mathbb K}\left(\frac{-k_\phi^2+\kappa ^2+1}{k_\phi^2+\kappa ^2+1}\right)\right]
\eeqa
where $k_\theta=\sqrt[4]{\gamma ^4-2 \gamma ^2+4 \gamma ^2 \ell_\theta^2+1}$ and
$k_\phi=\sqrt[4]{\kappa^4-2 \kappa^2+4 \gamma ^2 \ell_\phi^2+1}$. This set of equations solves the problem in a parametric form.
From the first four equations one should find four parameters $\ell_\theta,\ell_\phi,\kappa$ and $\gamma$
in terms of $\phi,\theta$ and $L$
and then the last equation gives the energy.

\subsection{Small angles}
In the small angle limit these equations simplify since then $\ell_\phi$ and $\ell_\theta$ are small
as is $\kappa-\gamma$. Still the solution is given in a parametric form in terms of a new parameter $\omega$ defined as
\beq\la{defOmega}
L\equiv 4g\[{\mathbb K}\left(\omega^2\right)-{\mathbb E}\left(\omega^2\right)\]\;.
\eeq
Let us first consider the case $\phi=0$ with $\theta$ small.

\paragraph{$\phi=0$ Case.}
In this case we see that $\ell_\phi=0$ and therefore $k_\phi=\sqrt{\kappa^2-1}$. The equations for $s$ and $E$ now reduce to
\beqa\la{s1app}
s&=&\frac{2}{\kappa}{\mathbb K}({1/\kappa^2})\\
\la{Eexpapp}
E&=&4g\[{\mathbb K}({1/\kappa^2})-{\mathbb E}({1/\kappa^2})\]\;.
\eeqa
Furthermore, if we expand the reminding equations to the leading order in $\ell_\theta$ we get
\beqa
s&=&\frac{2}{\gamma}{\mathbb K}({1/\gamma^2})\la{s2app}
-\ell_\theta^2 \[\frac{\left(\gamma ^3+\gamma \right)
   {\mathbb E}({1/\gamma^2})}{\left(\gamma ^2-1\right)^2}+\frac{\gamma  {\mathbb K}({1/\gamma^2})}{1-\gamma ^2}\]+{\cal O}(\ell_\theta^4)\\
\la{exLapp}
L&=&4 g \[{\mathbb K}({1/\gamma^2})-{\mathbb E}({1/\gamma^2})\]+\ell_\theta ^2 g
\[\frac{2  {\mathbb K}({1/\gamma^2})}{1-\gamma ^2}-\frac{4\gamma^2
   {\mathbb E}({1/\gamma^2})}{\left(\gamma ^2-1\right)^2}\]+{\cal O}(\ell_\theta^4)\\
\theta&=&\ell_\theta\frac{2\gamma {\mathbb E}({1/\gamma^2})}{\gamma^2-1}+{\cal O}(\ell_\theta^3)\;.\la{theex}
\eeqa
We want to express the result in terms of $\omega$ defined in \eq{defOmega}.
From \eq{exLapp} we find $\gamma$ in terms of $\omega$
\beq
\gamma=\frac{1}{\omega}+\ell_\theta^2\[\frac{1}{\omega(\omega^2-1)}+\frac{{\mathbb K}(\omega^2)}{2\omega{\mathbb E}(\omega^2)}\]\;.
\eeq
Then from \eq{s1app} and \eq{s2app} we find $\kappa$
\beq
\kappa=\frac{1}{\omega}-\frac{\ell_\theta^2}{2\omega}
\eeq
Substituting this into \eq{Eexpapp} and using \eq{theex} we get
\beq
E-L=2g\ell_\theta^2\frac{\omega^2{\mathbb E}(\omega^2)}{1-\omega^2}
=g\theta^2\frac{1-\omega^2}{2{\mathbb E}(\omega^2)}
\;.
\eeq
This is the result used in the main text.

\paragraph{General case.}
The procedure explained in the previous section can be applied to get expansion in both $\phi$
and $\theta$. We quote the result only
\beqa\
&&\frac{\nn E-L}{g (\phi^2-\theta^2)}\simeq -\frac{1-\omega ^2}{2 \mathbb{E}}
+
\theta ^2 \left(-\frac{  \mathbb{K}^2 \left(\omega ^2-1\right)^3}{32 \mathbb{E}^5 \omega
   ^2}+\frac{    \mathbb{K} \left(\omega ^2-9\right) \left(\omega ^2-1\right)^2}{96 \mathbb{E}^4
   \omega ^2}+\frac{    \left(\omega ^6-7 \omega ^2+6\right)}{96 \mathbb{E}^3 \omega
   ^2}\right)\\ \la{general}
   &&+\phi ^2 \left(-\frac{    \mathbb{K}^2 \left(\omega ^2-1\right)^3}{32 \mathbb{E}^5
   \omega ^2}-\frac{    \mathbb{K} \left(5 \omega ^2+3\right) \left(\omega ^2-1\right)^2}{96
   \mathbb{E}^4 \omega ^2}-\frac{    \left(5 \omega ^4-6 \omega ^2+1\right)}{96
   \mathbb{E}^3}\right) \\
\nn&&+\theta^4\(\frac{7   \mathbb{K}^4 \left(\omega ^2-1\right)^5}{768 \mathbb{E}^9 \omega ^4}-\frac{  \mathbb{K}^3
   \left(9 \omega ^2-37\right) \left(\omega ^2-1\right)^4}{768 \mathbb{E}^8 \omega ^4}+\frac{
   \mathbb{K}^2 \left(17 \omega ^4-138 \omega ^2+225\right) \left(\omega ^2-1\right)^3}{2304 \mathbb{E}^7
   \omega ^4}\right.\\
\nn&&\quad\quad\quad+\frac{  \mathbb{K} \left(2 \omega ^6+281 \omega ^4-1110 \omega ^2+1035\right) \left(\omega
   ^2-1\right)^2}{11520 \mathbb{E}^6 \omega ^4}\\
\nn&&\quad\quad\quad\left. +\frac{  (\omega^2 -1) \left(\omega ^8-16 \omega
   ^6+241 \omega ^4-570 \omega ^2+360\right)}{11520 \mathbb{E}^5 \omega ^4}\)\\
&&\nn+\phi^2\theta^2\(\frac{7   \mathbb{K}^4 \left(\omega ^2-1\right)^5}{768 \mathbb{E}^9 \omega ^4}-\frac{  \mathbb{K}^3
   \left(3 \omega ^2-17\right) \left(\omega ^2-1\right)^4}{384 \mathbb{E}^8 \omega ^4}-\frac{
   \mathbb{K}^2 \left(2 \omega ^4+27 \omega ^2-81\right) \left(\omega ^2-1\right)^3}{1152 \mathbb{E}^7
   \omega ^4}\right.\\
&&\nn\quad\quad\quad-\frac{  \mathbb{K} \left(14 \omega ^6-28 \omega ^4+135 \omega ^2-225\right) \left(\omega
   ^2-1\right)^2}{5760 \mathbb{E}^6 \omega ^4}\\
&&\nn\quad\quad\quad\left.+\frac{  (\omega^2 -1)  \left(-14 \omega ^8+14
   \omega ^6+31 \omega ^4-60 \omega ^2+45\right)}{11520 \mathbb{E}^5 \omega ^4}\)\\
&&+\phi^4\left(
\nn\frac{7  \mathbb{K}^4 \left(\omega ^2-1\right)^5}{768 \mathbb{E}^9 \omega ^4}+\frac{  \mathbb{K}^3
   \left(9 \omega ^2+19\right) \left(\omega ^2-1\right)^4}{768 \mathbb{E}^8 \omega ^4}+\frac{
   \mathbb{K}^2 \left(29 \omega ^4+30 \omega ^2+45\right) \left(\omega ^2-1\right)^3}{2304 \mathbb{E}^7
   \omega ^4}\right.\\
&&\nn\quad\quad\quad\left.+\frac{ \mathbb{K} \left(122 \omega ^6+11 \omega ^4+30 \omega ^2+45\right) \left(\omega
   ^2-1\right)^2}{11520 \mathbb{E}^6 \omega ^4}+\frac{ (\omega^2 -1) \left(61 \omega ^4-46
   \omega ^2+1\right)}{11520 \mathbb{E}^5}
\right)\;.
\eeqa
Here $\mathbb{K}\equiv \mathbb{K}(\omega^2)$ and similar for $\mathbb{E}$.
It is easy to get more orders in $\phi$ and $\theta$, however, the expressions become very bulky.
We found them explicitly up to the order $10$ and can provide them by request.

\paragraph{Small $L$.}
Further expanding at small $L$, we find
\beqa
\frac{ E-L}{\phi^2-\theta^2}&=&
 g\left(-\frac{1}{\pi }+\frac{ \theta ^2-5  \phi ^2 }{8 \pi
   ^3}+\frac{- \theta ^4+14  \phi ^2 \theta ^2-37 \phi ^4 }{64 \pi
   ^5}\right)\\
\nn&+&L  \left(\frac{3}{4 \pi ^2}-\frac{3 \left(\theta ^2-17 \phi
   ^2\right)  }{64 \pi ^4}+\frac{3 \left(5 \theta ^4-14 \phi ^2 \theta ^2+73 \phi ^4\right)
   }{256 \pi ^6}\right)\\
\nn&+&\frac{L^2}{g}  \left(-\frac{9}{64  \pi ^3}+\frac{3 \left(2
   \theta ^2-9 \phi ^2\right)   }{128  \pi ^5}-\frac{3 \left(13 \theta ^4-138 \phi ^2 \theta
   ^2+653 \phi ^4\right)   }{8192  \pi ^7}\right)\\
\nn&+&\frac{L^3}{g^2}  \left(-\frac{5}{256
   \pi ^4}+\frac{-133 \theta ^2-379 \phi ^2 }{8192  \pi
   ^6}+\frac{-41 \theta ^4-2 \phi ^2 \theta ^2-341 \phi ^4}{4096 \pi
   ^8}\right)\\
\nn&+&\frac{L^4}{g^3}  \left(\frac{45}{16384 \pi ^5}-\frac{105 \left(11 \theta ^2-7 \phi ^2\right)   }{131072
    \pi ^7}-\frac{3 \left(2001 \theta ^4+3698 \phi ^2 \theta ^2-139 \phi ^4\right)
   }{1048576  \pi ^9}\right)
\eeqa
The first line reproduces the known result of \cite{Drukker:2011za}.
For $\theta\simeq\phi$ we get
\beqa
\la{ELex}\frac{ E-L}{\phi^2-\theta^2}&=&
\left(-\frac{g}{\pi }+\frac{3 L}{4 \pi ^2}-\frac{9 L^2}{64
   g\pi  }-\frac{5 L^3}{256  g^2 \pi  }+\frac{45 L^4}{16384 g^3 \pi
   ^5}\right)\\
\nn&+&\phi ^2
    \left(-\frac{g}{2 \pi ^3}+\frac{3 L}{4 \pi ^4}-\frac{21 L^2}{128 g  \pi  }-\frac{L^3}{16
    g^2 \pi  }-\frac{105 L^4}{32768  g^3 \pi
    }\right)\\
\nn&+&\phi ^4  \left(-\frac{3 g}{8 \pi
   ^5}+\frac{3 L}{4 \pi ^6}-\frac{99 L^2}{512   g\pi  }-\frac{3 L^3}{32  g^2 \pi
    }-\frac{2085 L^4}{131072  g^3 \pi  }\right)\\
\nn&+&\phi ^6  \left(-\frac{5 g}{16 \pi ^7}+\frac{3 L}{4 \pi
   ^8}-\frac{225 L^2}{1024  g \pi  }-\frac{L^3}{8  g^2 \pi  }-\frac{7905
   L^4}{262144  g^3 \pi  }\right)\\
\nn&+&\phi ^8  \left(-\frac{35 g}{128 \pi ^9}+\frac{3 L}{4 \pi ^{10}}-\frac{1995 L^2}{8192 g \pi^{11}}-\frac{5 L^3}{32  g^2 \pi  }-\frac{97425 L^4}{2097152  g^3 \pi
    }\right)\;.
\eeqa

\paragraph{Large $L$.}
For $L/g$ to be large, $\omega$ should be exponentially close to $1$
\beqa
\omega&=&1+
8 \delta+\delta ^2 \left(32-\frac{16 L}{g}\right) +\delta ^3 \left(\frac{48 L^2}{g^2}-\frac{112 L}{g}+224\right)\\
&+&\delta ^4 \left(-\frac{512 L^3}{3
   g^3}+\frac{384 L^2}{g^2}-\frac{1344 L}{g}+1280\right)+{\cal O}(\delta^5)\nn
\eeqa
where $\delta\equiv e^{-\frac{L}{2g}-2}$. One can plug this expansion of $\omega$ into the expansions \eq{general} to get the large $L$ expansion.
We do not use this limit in the paper so we skip the details.

\section{Strong Coupling Expansion for Fixed $L$}\la{strogapp}
Here we collect the strong coupling expansion of our result for various $L$'s.
\beq
\begin{array}{l|l}
L& B_L+{\cal O}(1/g^4)\\ \hline
 0 & \frac{g}{\pi }-\frac{3}{8 \pi ^2}+\frac{3}{128 \pi ^3 g}+\frac{3}{512 \pi ^4 g^2}+\frac{63}{32768 \pi ^5
   g^3} \\
 1 & \frac{g}{\pi }-\frac{9}{8 \pi ^2}+\frac{39}{128 \pi ^3 g}+\frac{39}{512 \pi ^4 g^2}+\frac{279}{32768 \pi ^5
   g^3}\\
 2 & \frac{g}{\pi }-\frac{15}{8 \pi ^2}+\frac{111}{128 \pi ^3 g}+\frac{165}{512 \pi ^4 g^2}-\frac{1449}{32768 \pi
   ^5 g^3}\\
 3 & \frac{g}{\pi }-\frac{21}{8 \pi ^2}+\frac{219}{128 \pi ^3 g}+\frac{441}{512 \pi ^4 g^2}-\frac{9441}{32768 \pi
   ^5 g^3}\\
 4 & \frac{g}{\pi }-\frac{27}{8 \pi ^2}+\frac{363}{128 \pi ^3 g}+\frac{927}{512 \pi ^4 g^2}-\frac{30177}{32768 \pi
   ^5 g^3} \\
 5 & \frac{g}{\pi }-\frac{33}{8 \pi ^2}+\frac{543}{128 \pi ^3 g}+\frac{1683}{512 \pi ^4 g^2}-\frac{72297}{32768
   \pi ^5 g^3}
\end{array}
\eeq
This table can be summarized by the following single equation
\beqa
\nn B_L&=&\frac{g}{\pi }-\frac{6 L+3}{8 \pi ^2}+\frac{3 \left(6 L^2+6 L+1\right)}{128 \pi ^3 g}+\frac{10 L^3+15 L^2+11 L+3}{512 \pi ^4 g^2}\\
   &-&\frac{9 \left(10 L^4+20 L^3-22 L^2-32 L-7\right)}{32768 \pi ^5 g^3}+{\cal O}\left(1/g^4\right)\;.
\eeqa
We see that knowing just a few first $L$'s we can in fact extrapolate to large $L$ using the polynomial property of the coefficients.
In particular we can make a comparison with the classical limit by re-expanding for large $L\sim g$ \`a la \cite{Roiban:2011fe}. Denoting ${\mathfrak L}\equiv L/g$ we get
\beqa
B_L &=&g \left(\frac{1}{\pi }-\frac{3 \mathfrak{L}}{4 \pi ^2}+\frac{9 \mathfrak{L}^2}{64 \pi ^3}+\frac{5
   \mathfrak{L}^3}{256 \pi ^4}-\frac{45 \mathfrak{L}^4}{16384 \pi
   ^5}+{\cal O}\left(\mathfrak{L}^{5}\right)\right)\\
\nn   &+&g^0\left(-\frac{3}{8 \pi ^2}+\frac{9 \mathfrak{L}}{64 \pi ^3}+\frac{15
   \mathfrak{L}^2}{512 \pi ^4}-\frac{45 \mathfrak{L}^3}{8192 \pi
   ^5}+{\cal O}\left(\mathfrak{L}^{4}\right)\right)\\
\nn   &+&\frac{1}{g}\(\frac{3}{128 \pi ^3}+\frac{11 \mathfrak{L}}{512 \pi
   ^4}+\frac{99 \mathfrak{L}^2}{16384 \pi ^5}+{\cal O}\left(\mathfrak{L}^{3}\right)\)\\
\nn   &+&\frac{1}{g^2}\({\frac{3}{512 \pi
   ^4}+\frac{9 \mathfrak{L}}{1024 \pi ^5}+{\cal O}\left(\mathfrak{L}^{2}\right)}\)
\nn   +\frac{1}{g^3}\(\frac{63}{32768 \pi ^5}+{\cal O}\left(\mathfrak{L}^{1}\right)\)+{\cal O}(1/g^4)\;
\eeqa
we see that the first line by dimension is the classical contribution. The second line is the one-loop part of the result and so on.
Indeed, this procedure gives the correct classical energy in the first line of \eq{ELex}! This is another more direct test of our results
which does not involve the algebraic curve consideration. Note that the first term in the second line in the above equation
perfectly reproduces the one loop result of \cite{Drukker:2011za}.

\end{document}